\documentclass[a4paper,11pt]{article}
\usepackage{jheppub}\usepackage{lineno}
\usepackage{amsthm,amsmath,amssymb,amsfonts}
\usepackage{mathrsfs,bm}\usepackage{multirow}
\usepackage{slashed,braket,tensor,cancel}
\usepackage[dvipsnames]{xcolor}
\usepackage{subfig} 
\usepackage{tikz,tikz-cd,tkz-euclide}
\usepackage[compat=1.1.0]{tikz-feynman}
\usetikzlibrary{cd}\usetikzlibrary{calc}\usetikzlibrary{patterns}
\usetikzlibrary{feynman}\tikzfeynmanset{warn luatex=false}\usetikzlibrary{arrows.meta}
\usetikzlibrary{decorations.markings}
\tikzset{mid arrow/.style={
postaction={decorate,decoration={markings,mark=at position 0.5 with {\arrow{>}}}}}}
\usetikzlibrary{decorations.pathmorphing}
\tikzset{snake it/.style={decorate, decoration=snake}}

\allowdisplaybreaks[4]

\newcommand{\bea}{\begin{align}}\newcommand{\eea}{\end{align}}
\newcommand{\be}{\begin{align}}\newcommand{\ee}{\end{align}}
\def\eq{\eqref}\def\nn{\nonumber}

\def\c.c.{\mathrm{c.c.}}\def\hc{\mathrm{h.c.}}

\def\d{\mathrm{d}}\def\i{\mathrm{i}}
\newcommand{\e}{\mathrm{e}}\newcommand{\p}{\partial} 

\def\0{{(0)}}\def\1{{(1)}}\def\2{{(2)}}\def\3{{(3)}}\def\4{{(4)}}
\newcommand{\Rn}[1]{{\rm\uppercase\expandafter{\romannumeral#1}}}

\def\qrq{\quad\Rightarrow\quad}
\def\qlq{\quad\Leftrightarrow\quad}
\def\qaq{\quad\text{and}\quad}

\def\qwq{\quad\text{with}\quad}

\def\wt{\widetilde}

\newcommand{\SO}{\mathrm{SO}}
\newcommand{\so}{\mathfrak{so}}

\def\ab{{\alpha\beta}}\def\mn{{\mu\nu}}\def\rs{{\rho\sigma}}\def\nm{{\nu\mu}}
\def\a{\alpha}\def\b{\beta}\def\g{\gamma}
\def\ep{\epsilon}\def\th{\theta}
\def\k{\kappa}\def\l{\lambda}
\def\m{\mu}\def\n{\nu}
\def\r{\rho}\def\s{\sigma}\def\t{\tau}
\def\vp{\varphi}
\def\Th{\Theta}\def\L{\Lambda}

\def\ca{\mathcal{A}}

\def\cf{\mathcal{F}}
\def\ch{\mathcal{H}}
\def\ci{\mathcal{I}}
\def\cl{\mathcal{L}}
\def\cm{\mathcal{M}}
\def\co{\mathcal{O}}
\def\cs{\mathcal{S}}\def\ct{\mathcal{T}}

\def\cA{\mathcal{A}}

\def\cF{\mathcal{F}}
\def\cH{\mathcal{H}}

\def\cM{\mathcal{M}}

\def\cS{\mathcal{S}}

\def\R{\mathbb{R}}

\arxivnumber{2508.15619} 
\title{\boldmath Extrapolating the massive fields to future timelike infinity}
\author{Wen-Bin Liu}\author{and Jiang Long}
\affiliation{School of Physics, Hua-Zhong University of Science and Technology,\\
Luoyu Road 1037, Wuhan, Hubei 430074, China}
\emailAdd{liuwenbin0036@hust.edu.cn, longjiang@hust.edu.cn}
\abstract{It is well-known that future timelike infinity ($i^+$) in four-dimensional Minkowski spacetime is conformal to the unit three-dimensional hyperboloid ($H^3$). We asymptotically expand massive fields with spin $0,1,2$ near $i^+$ and extrapolate them onto this hyperboloid. These fields oscillate with a frequency equal to their mass and exhibit a universal asymptotic decay $\tau^{-3/2}$. The fundamental fields are free and encode the outgoing scattering data. They are local operators defined on the boundary $H^3$ with which we construct the Poincar\'e charges. The Poincaré algebra can be extended to $\text{MDiff}(H^3)\ltimes C^{\infty}(H^3)$ using smeared operators associated with energy and angular momentum densities. For spinning fields, a spin operator must be included to close the algebra. The extended algebra shares the same form as the five-dimensional intertwined Carrollian diffeomorphism and reduces to the BMS algebra at $i^+$ by restricting the choice of test functions and vectors.}

\begin{document}
\maketitle
\flushbottom
\section{Introduction}
Recently, we have constructed a series of flux algebras by performing asymptotic expansion and quantization of massless fields at future/past null infinity $\mathcal I^{\pm}$ \cite{Liu:2022mne,Liu:2023qtr,Liu:2023gwa,Li:2023xrr,Liu:2023jnc,Liu:2024nkc,Liu:2024rvz,Guo:2024qzv}. In particular, a new operator called the helicity flux operator emerges which concerns (electromagnetic) duality transformation for the spinning field. The helicity flux operator measures the difference in the number of particles with opposite helicities and is also related to the second Chern character and the chiral anomaly \cite{Nakahara:2003nw}. In higher $d$ dimensions ($d>4$), there exist multiple helicity directions \cite{Liu:2024rvz}, since the little group for massless particles becomes $\SO(d-2)$ rather than a simple phase transformation. 

Interestingly, the flux algebras contain the BMS algebra as a sub-algebra that governs the infrared physics at null infinity and has attracted considerable attention over the last decade since the discovery of the infrared triangle \cite{Strominger:2013jfa,He:2014laa,Strominger:2014pwa,Strominger:2017zoo}. The BMS group \cite{bms1,bms2} has been identified with the conformal Carroll group of level 2 \cite{Duval_2014a,Duval_2014b,Duval:2014uoa}, which originates from the fact that null infinity is a Carrollian manifold \cite{Une,Gupta1966OnAA}. Moreover, the standard BMS group has been enhanced to the extended BMS group \cite{Barnich:2009se,Barnich:2010eb,Barnich:2010ojg,Barnich:2011mi} which admits the superrotation generated by a local conformal Killing vector (CKV) on the celestial sphere. One can also allow the boundary structure to fluctuate such that the Lorentz transformation can be enlarged to a general diffeomorphism on the celestial sphere, and one obtains the generalized BMS group \cite{Campiglia:2014yka,Campiglia:2015yka,Compere:2018ylh,Campiglia:2020qvc}. 

Although the above asymptotic symmetries were originally developed in the context of asymptotically flat gravity, they can also be realized in the quantum field theory. Once fixing the spacetime background to the Minkowski metric, it is natural to consider various matter fields, including both bosons and fermions. More precisely, the supertranslation and superrotation can be realized through quantum flux operators. These quantum fluxes are derived from
\begin{align}
\mathscr F_{f,Y}=\int_{\ci^+}(\d^3 x)^\m T_\mn\xi_{f,Y}^\n,
\end{align}
where $T_\mn$ is the matter stress tensor and $\xi_{f,Y}$ denotes the generators for supertranslation and superrotation.\footnote{This procedure is well-known for the Killing vector, i.e., the Poincar\'e generator (also a subset of supertranslation and superrotation generators). Equivalently, extracting the energy and angular momentum densities at $\ci^+$ and then integrating them with appropriate parameters will lead to the same result.} They form our flux algebra under the quantum commutator. In the framework of bulk reduction, we reduce a well-known bulk quantum field theory to $\ci^+$ through the large $r$ expansion (while keeping the retarded time $u$ unchanged). We use $\cf$ to represent the leading order field in the expansion of bulk field $F$ (e.g., $A_A$ for bulk field $a_A$ in \cite{Liu:2023qtr}). We call $\cf$ the fundamental field or boundary field, and it encodes the radiative degree of freedom. It is unconstrained, and the subleading fields can be solved from the bulk equation of motion in terms of $\cf$.

The quantum flux $\mathscr F_{f,Y}$ should generate the corresponding transformation when it acts on the fundamental field $\cf$
\begin{align}
[\i\mathscr F_{f,Y},\cf]=\delta_{f,Y}\cf,
\end{align}
where the classical variation $\delta_{f,Y}\cf$ is induced from the bulk Lie derivative $\cl_{\xi_{f,Y}}F$. However, this only holds for the scalar field under $\xi_{f,Y}$ and the spinning field under the supertranslation. That is because the Minkowski metric changes under superrotation, and so does the boundary metric $\g_{AB}$. The non-vanishing $\delta_Y\g_{AB}$ violates the underlying requirement for a quantum field theory, and it will have an effect when computing the commutator for the spinning field under superrotation since $\mathscr F_Y$ is made up of not only the fundamental field but also the boundary metric. One should modify the classical variation $\delta_Y\cf$ to the covariant variation $\Delta_Y\cf$ such that it can match the commutator
\begin{align}
[\i\mathscr F_Y,\cf]=\Delta_Y\cf,\label{agreement}
\end{align}
The so-called covariant variation is constructed for an arbitrary boundary tensor field $V_{A_1\cdots A_n}$ as follows
\begin{align}
\Delta_YV_{A_1\cdots A_n}=\delta_YV_{A_1\cdots A_n}-\frac{1}{2}\sum_{i=1}^n\delta_Y\g_{A_iB}V_{A_1\cdots A_{i-1}}{}^B{}_{A_{i+1}\cdots A_n}
\end{align}
so that it preserves the boundary metric $\Delta_Y\g_{AB}=0$ and thus is adapted to the boundary theory. This property makes sure the agreement \eq{agreement}. \emph{The above is the overall logic of our previous papers.}



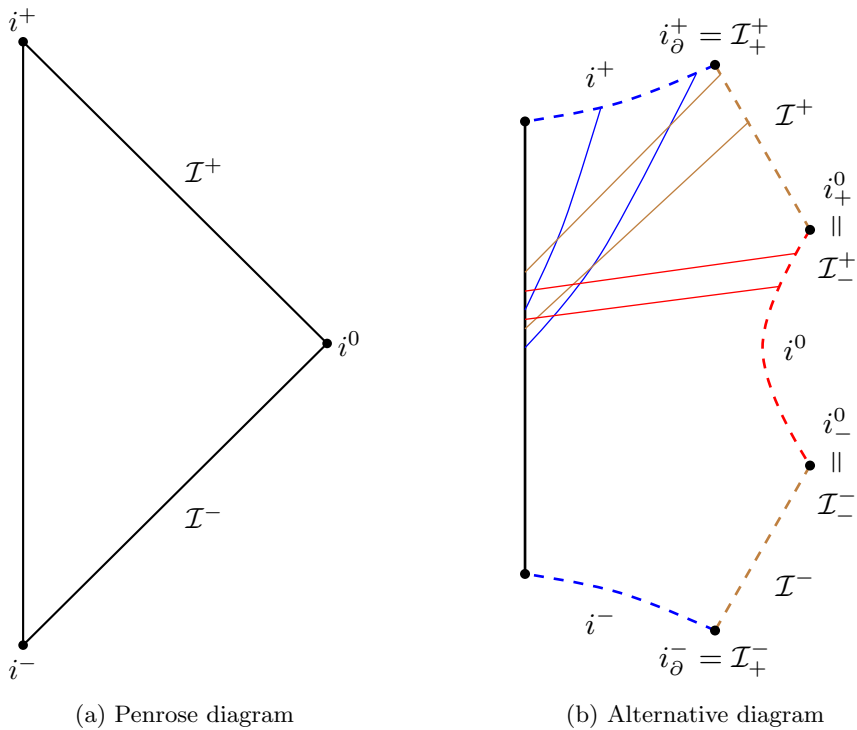
\begin{figure}[tbp]
\centering
\subfloat[Penrose diagram]{
\begin{tikzpicture}
\draw[thick] (0,4) node[above]{$i^+$} -- (2,2) node[above right]{$\mathcal{I}^+$} -- (4,0) node[right]{$i^0$}  -- (2,-2) node[below right]{$\mathcal{I}^-$} -- (0,-4) node[below]{$i^-$} -- cycle;
\fill (4,0) circle (1.8pt);
\fill (0,4) circle (1.8pt);
\fill (0,-4) circle (1.8pt);
\end{tikzpicture}
}\hfil
\subfloat[Alternative diagram]{\scalebox{1.25}{
\begin{tikzpicture}
\draw[thick,blue,dashed] plot[smooth,tension=0.7] coordinates{(2,3) (1,2.6) (0,2.4)};
\draw[thick,blue,dashed] plot[smooth,tension=0.7] coordinates{(2,-3) (1,-2.6) (0,-2.4)};
\draw[thick,red,dashed] plot[smooth,tension=0.7] coordinates{(3,1.25) (2.5,0) (3,-1.25)};
\draw[thick,brown,dashed] (3,1.25)--(2,3);
\draw[thick,brown,dashed] (3,-1.25)--(2,-3);
\draw[thick] (0,-2.4)--(0,2.4);
\fill (2,3) circle (1.5pt);
\fill (2,-3) circle (1.5pt);
\fill (0,2.4) circle (1.5pt);
\fill (0,-2.4) circle (1.5pt);
\fill (3,1.25) circle (1.5pt);
\fill (3,-1.25) circle (1.5pt);
\node at (0.8,2.9) {\footnotesize $i^+$};
\node at (2.85,2.5) {\footnotesize $\ci^+$};
\node at (2.85,-2.5) {\footnotesize $\ci^-$};
\node at (2.85,0) {\footnotesize $i^0$};
\node at (0.8,-2.9) {\footnotesize $i^-$};
\node at (2,3.3) {\footnotesize $i^+_\p=\ci^+_+$};
\node at (2,-3.3) {\footnotesize $i^-_\p=\ci^-_-$};
\node at (3.3,1.25) {\footnotesize $\begin{matrix}
i^0_+\\ \rotatebox{90}{=}\\ \ci^+_-
\end{matrix}$};
\node at (3.3,-1.25) {\footnotesize $\begin{matrix}
i^0_-\\ \rotatebox{90}{=}\\ \ci^-_+
\end{matrix}$};
\draw[blue] plot[smooth,tension=0.6] coordinates{(0,0) (0.8,1) (1.8,2.9)};
\draw[blue] plot[smooth,tension=0.6] coordinates{(0,0.4) (0.4,1.3) (0.8,2.55)};
\draw[brown] (0,0.2)--(2.36,2.4);
\draw[brown] (0,0.8)--(2.06,2.9);
\draw[red] (0,0.3)--(2.68,0.65);
\draw[red] (0,0.6)--(2.85,1);
\end{tikzpicture}
}}
\caption{Two diagrams for the asymptotically Minkowski spacetime. The left one is the standard Penrose diagram, while the right one is schematically an alternative description. In the latter diagram, more structure can be depicted, such as the joint corners of different asymptotic regions. Moreover, two geodesics from bulk to $i^+/\ci^+$ are drawn with the colors of blue/brown which describe massive/massless particles, while the red geodesics approach $i^0$. Note that the right diagram is similar to the one in \cite{Compere:2023qoa} except that we keep the null infinity represented by an oblique line.}
\label{fig:placeholder}
\end{figure}

The same logic can be applied to massive fields. Massive particles depart from/arrive at past/future timelike infinity ($i^\mp$), which is also part of the conformal boundary of an asymptotically flat spacetime.  As depicted in figure \ref{fig:placeholder}, the conformal boundary of an asymptotically Minkowski spacetime is divided into three parts according to the category of the approaching geodesics. Among them, the timelike infinity and spatial infinity $i^0$ are conformal to the unit hyperboloid\footnote{One should be careful with the  signature of the boundary manifold. }, while the null infinity has the topology of $\R\times S^2$. Timelike infinity has a dual description with the spatial infinity at which one can also define the BMS algebra \cite{Troessaert:2017jcm,Henneaux:2018cst,Henneaux:2018hdj}. It was also found that one could also study the asymptotic symmetry \cite{Cutler1989PropertiesOS,porrill_1982,Gen:1997az} and even define supertranslation \cite{Chakraborty:2021sbc,Compere:2023qoa} near timelike infinity. Moreover, one can lift the timelike infinity to a Carrollian manifold called Ti just as blowing up the spatial infinity to Spi \cite{Ashtekar:1978zz,Gibbons:2019zfs,Figueroa-OFarrill:2021sxz}. The massive Carrollian field on Ti has been investigated in \cite{Have:2024dff}. Early efforts on the massive field near timelike infinity of Minkowski spacetime can be found in \cite{Campiglia:2015qka,Campiglia:2015kxa}. 

One of the main motivations of this work is to extrapolate the massive bulk field to $i^+$, realize the Poincar\'e algebra using the boundary field, and extend the Poincar\'e algebra in the framework of bulk reduction \cite{Liu:2024nkc}. Another motivation is to complete the picture of flat space holography \cite{Susskind:1998vk,Giddings:1999jq,deBoer:2003vf,Arcioni:2003xx}. The flat space holography aims at applying the holographic principle \cite{tHooft:1993dmi,Susskind:1994vu}, which has achieved great success in the AdS/CFT correspondence \cite{Maldacena:1997re,Witten:1998qj,Aharony:1999ti}, to the more realistic asymptotically flat spacetime. Two approaches called Carrollian and celestial holographies have been proposed \cite{Bagchi:2016bcd,Ciambelli:2018wre,Pasterski:2016qvg,Pasterski:2021raf,Donnay:2022aba,Bagchi:2022emh,Donnay:2022wvx,Donnay:2023mrd}. However, the Carrollian approach is restricted to massless scattering, which is of course important, but not the only physically interesting thing. Massive scattering also matters, especially when the bulk matter field is considered.

In the following, we exhibit the main results of this paper. At first, the timelike infinity $i^+$ is conformal to a hyperboloid $H^3$, which suffices to encode the scattering data of massive fields. A bulk field $F$ exhibits universal decay 
\begin{align}
F=\frac{\e^{-\i m\tau}}{\tau^{3/2}}(\cF+\cdots)+\frac{\e^{\i m\tau}}{\tau^{3/2}}(\cF^\dagger+\cdots)
\end{align} 
near $i^+$ under the Cartesian frame where $\tau$ is the time in the hyperboloid coordinates \eq{taurhodef} and $m$ is the mass of the field. The leading spatial components $\mathcal F$ and $\mathcal F^\dagger$ in the hyperboloid frame are understood as the fundamental fields and correspond to annihilation and creation operators in the canonical quantization, respectively. 
Second, we construct the Poincar\'e charges as quadratic composite operators of the fundamental fields. The Lorentz transformation is mapped to the isometry group of $H^3$ and thus is generated by the Killing vector \eqref{Xmna}. On the other hand, the global space and time translations are mapped to phase rotations that are parameterized by a single function which obeys equation \eqref{transfh}. Third, we extend the Killing vector to any smooth and divergence-free vector $X^a$ and extend the phase function to any smooth function $f$ on $H^3$ and then construct the associated charge operators. The extended algebra contains the BMS algebra as a sub-algebra at $i^+$. Note that the realization of BMS algebra using a massive scalar field can be found in \cite{Longhi:1997zt}. The extended charge operators are not conserved in general. However, their actions on the fundamental field could match those from a certain bulk Lie derivative. Similar to the case at null infinity, one should lift the Lie derivative to covariant variation for any spinning field, and this leads to a new operator. At last, we verify that this operator is a spin operator which characterizes the spin density at $i^+$ of the massive field. It is also checked in various ways that the spin operator indeed reduces to spin angular momentum, including converting to a locally flat frame, performing the mode expansion, and using Noether's procedure.

The layout of this paper is as follows. In section \ref{poin}, we will discuss how the Poincar\'e algebra is realized at timelike infinity via massive fields with spin $0,1$, and 2. Conventions and notations are also established. In the following section, we extend the Poincar\'e charges and compute the corresponding algebras. In section \ref{sd}, we discuss various aspects of the spin density operator emerging from the algebra for the spinning field. In section \ref{comp}, the extended algebra is reduced to the BMS algebra {and our result is compared with the literature}. We will summarize the work in the last section and discuss future directions that deserve study. We briefly display the extended charge algebra in general dimensions and particularly in 3 dimensions in appendix \ref{secddim}.

\section{Realizing Poincar\'e algebra through massive fields}\label{poin}
\subsection{Massive scalar}
As a warm-up, we will define the Poincar\'e charges at future timelike infinity $i^+$ for the massive scalar theory in this subsection. The conventions and notations are established simultaneously. 
\subsubsection{Background spacetime}
To describe future timelike infinity $i^+$, one introduces the hyperbolic coordinates \cite{Campiglia:2015qka,Campiglia:2015kxa,Chakraborty:2021sbc,Compere:2023qoa}
\begin{align}
t=\tau\cosh\rho,\quad r=\tau\sinh\rho \qlq \t=\sqrt{t^2-r^2},\quad \r={\rm arctanh}\frac{r}{t}\label{taurhodef}
\end{align}
such that the Minkowski metric takes the form
\begin{align}
\mathrm{d} s^2=-\mathrm{d} \tau^2+\tau^2h_{ab}\mathrm{d} y^a\mathrm{d} y^b\equiv \eta_\ab\d x^\a\d x^\b,\label{habmet}
\end{align}
in terms of $x^\a=(\t,\r,\theta,\phi)$. We denote $x^A=(\th,\phi)$ which is also written as $\Omega$ for convenience, while $y^a=(\r,\theta,\phi)$ covers the unit hyperboloid $H^3$ whose metric $h_{ab}$ reads 
\begin{align}
h_{ab}\mathrm{d} y^a\mathrm{d} y^b=\mathrm{d} \rho^2+\sinh^2\rho\gamma_{AB}\mathrm{d} x^A\mathrm{d} x^B
\end{align}
which is the $\t\to\infty$ limit of \eq{habmet} after Weyl scaling and thus also describes $i^+$. The metric of the unit sphere is denoted by $\gamma_{AB}$, whose components are
\begin{align}
\gamma_{AB}=\left(\begin{array}{cc}1&0\\0&\sin^2\theta\end{array}\right).
\end{align}

One can easily compute the non-vanishing components of the Christoffel symbol
\begin{subequations}
\begin{align}
\Gamma^\t_{ab}={\t}h_{ab},\qquad \Gamma^a_{\t b}=\frac{1}{\t}\delta^a_{b},\qquad \Gamma^c_{ab}=\Gamma^c_{ab}[h],\label{ksf}
\end{align}
where $\Gamma^c_{ab}[h]$ are those for $h_{ab}$ whose non-vanishing components are
\begin{align}
&\Gamma^\r_{AB}=-\cosh\r\sinh\r\g_{AB},\qquad \Gamma^\th_{\th\r}=\Gamma^\phi_{\phi\r}=\coth\r,\\
&\Gamma^\theta_{\phi\phi}=-\cos\theta\sin\theta,\qquad \Gamma^\phi_{\theta\phi}=\cot\theta.
\end{align}
\end{subequations}
$H^3$ is a maximally symmetric space whose Riemann tensor reads 
\begin{align}
    R_{abcd}=\frac{1}{6}(h_{ac}h_{bd}-h_{ad}h_{bc})R\qwq R=-6.
\end{align}
For a constant $\tau$ hypersurface $\mathcal H_\tau$, the normal vector is $\partial_\tau$ and the associated normal covector is $-\d\tau$.
The volume form of $\mathcal H_\tau$ reads
\begin{align}
(\d^3x)_\t=-\t^3\sinh^2\r\sin\th\d\r\wedge\d\th\wedge\d\phi
\end{align}
whose integral is denoted as
\begin{align}
\int_{\mathcal H_\tau}(\d^3x)_\t (\cdots)^\t\equiv-\t^3\int\d^3y \sqrt{h}\, (\cdots)^\t
\end{align} where the integration 
\begin{align}
\int \d^3y \sqrt{h}\,=\int \sinh^2\rho \sin\theta \d\rho \d\theta \d\phi 
\end{align} 
is the one on $H^3$.

Using $x^\m=(t,x^i)$ to denote the Cartesian coordinates, one can calculate the Jacobian matrices
\begin{subequations}\label{Jacobi}
\begin{align}
\frac{\p x^\a}{\p x^\m}&=t_\m\delta^\a_\t-\frac{1}{\t}\nabla^at_\m\delta^\a_a,\\ 
\frac{\p x^\m}{\p x^\a}&=-t^\m\delta^\t_\a-\t \nabla_at^\m\delta^a_\a.
\end{align}
\end{subequations}
Here, we have defined
\begin{align}
t_\m\equiv\frac{\p\t}{\p x^\m}=-\frac{1}{\t}x_\m=(\cosh\r,-n_i\sinh\r) \label{tmu}
\end{align}
with $n^i$ the unit normal vector of the sphere and then $t^\m=-\p_\t x^\m$. The vector $t_\mu$ is related to the normal covector $\d\tau$ via 
\begin{align}
\d\tau=t_\mu \d x^\mu.
\end{align} 
Note that the covariant derivatives $\nabla_\a$, $\nabla_a$, and $\nabla_A$ are adapted to $\eta_\ab$, $h_{ab}$, and $\g_{AB}$, respectively.

More explicitly, we define
\begin{align}
S_\m^a\equiv -\nabla^at_\m=(s_\m,-Y^A_\m\sinh^{-1}\r )
\end{align}
with
\begin{align}
s_\m=-\p_\r t_\m=-(\sinh\r, n_i\cosh\r),
\end{align}
and $Y^A_\m=-\nabla^An_\m$ with $n_\m=(-1,n_i)$. In reverse, we have $t_\m=-\p_\r s_\m$. It is easy to verify the following identities
\begin{align}
S_\m^aS^\m_b=\delta^a_b,\qquad S^\m_aS^a_\n =s^\m s_\n+\g^\m_\n,
\end{align}
which ensure the consistency relations for Jacobian matrices
\begin{subequations}
\begin{align}
\frac{\p x^\a}{\p x^\m}\frac{\p x^\m}{\p x^\b}&=\delta^\a_\t\delta^\t_\b+\delta^\a_\r\delta^\r_\b+\delta^\a_A\delta^A_\b=\delta^\a_\b,\\
\frac{\p x^\m}{\p x^\a}\frac{\p x^\a}{\p x^\n}&=-t^\m t_\n+s^\m s_\n+\g^\m_\n=\delta^\m_\n.
\end{align}
\end{subequations}
Here, the symmetric tensor $\gamma_{\mu\nu}$ is the bulk version of the metric $\g_{AB}$ on the sphere, and it is related to the CKVs on $S^2$ through 
\begin{align}
\gamma_{\mu\nu}=\g_{AB}Y_\mu^A Y_{\nu}^B.
\end{align}
Moreover, we have
\begin{align}
s_\m s^\m=-t_\m t^\m=1,\quad t_\m s^\m=0,\quad s^\m Y^{A}_\m=t^\m Y^{A}_\m=0,
\end{align}
and
\begin{align}
t_\m=\p_\r^2 t_\m, \quad s_\m=\p_\r^2 s_\m.
\end{align}

One can write the translation generator $\p_\m$ as
\begin{align}
\p_\m=t_\m\p_\t+\frac{1}{\t}S^a_\m\p_a.\label{tipm}
\end{align}
When acting on a scalar, we have
\begin{align}
\Box\equiv\p_\m\p^\m=\eta^\ab (\p_\a\p_\b-\Gamma_{\ab}^\g \p_\g)
\end{align}
which takes the form
\begin{align}
\Box=-\p_\t^2-\frac{3}{\t}\p_\t+\frac{1}{\t^2} \nabla_a\nabla^a\label{lpls}
\end{align}
with
\begin{align}
\nabla_a\nabla^a=\p_\r^2+2\coth\r\p_\r+\frac{1}{\sinh^2\r}\nabla_A\nabla^A.
\end{align}

It is easy to verify 
\begin{align}
(\nabla_a\nabla^a-3) t_\m=0
\end{align}
where the differential operator on the left is the trace of $\nabla_a\nabla_b-h_{ab}$. In fact, $t_\m$ encodes four independent solutions to
\begin{align}
(\nabla_a\nabla_b-h_{ab})f=0\label{transfh}
\end{align}
which is one of the characterizations of translation generator. 

As for the Lorentz transformation, one can compute
\begin{align}
x_\m\p_\n-x_\n\p_\m&=(t_\m\p_\r t_\n-t_\n\p_\r t_\m)\p_\r+\frac{1}{\sinh^2\r}(t_\m\nabla^A t_\n-t_\n\nabla^A t_\m)\p_A\equiv X^a_\mn\p_a,
\end{align}
where we have defined
\begin{align}
X^a_\mn=t_\m \nabla^a t_\n-t_\n \nabla^a t_\m\label{Xmna}
\end{align}
whose components satisfy
\begin{align}
X^\r_\mn=-\frac12\tanh\r\nabla_AX_\mn^A=
\begin{pmatrix}
0&-n_i\\ n_i&0
\end{pmatrix}.
\end{align}
It is easy to check that $X^a_\mn$ solves the Killing equation on $H^3$.

\subsubsection{Massive scalar and its asymptotic expansion}
The Lagrangian for a massive scalar $\Phi$ with mass $m$ takes the following form
\begin{align}
\cl=-\frac{1}{2}\p_\m\Phi\p^\m\Phi-\frac{1}{2}m^2\Phi^2-V[\Phi],\label{scalarL}
\end{align}
which results in the Klein-Gordon equation
\begin{align}
(\Box-m^2)\Phi-V'[\Phi]=0.
\end{align} 
The potential term $V[\Phi]$ may be expanded around $\Phi=0$ perturbatively 
\begin{align}
V[\Phi]=\lambda \Phi^{k}+\cdots
\end{align}
with $k>2$.

Near $i^+$, one can use the saddle-point approximation to reduce the bulk mode expansion for the massive scalar
\begin{align}
\Phi(x)=\int \frac{\d^3p}{(2\pi)^3}\frac{1}{2\omega_{\bm p}}[a(\bm p)\e^{\i p\cdot x}+\hc].
\end{align}
More explicitly, we parameterize the position and momentum as
\begin{align}
x^\m=\t(\cosh\r,n^i\sinh\r)=-\t t^\m,\qquad p^\m=m\hat p^\m=m(\sqrt{1+\hat{p}_i^2},\hat{p}^i).
\end{align}
As a consequence, the plane wave $\e^{\i p\cdot x}$ becomes
\begin{align}
\e^{\i p\cdot x}=\e^{\i m\t\zeta(\hat p)} \qwq \zeta(\hat p)\equiv-t\cdot\hat p=-\sqrt{1+\hat{p}_i^2}\cosh\r+n_i\hat p^i\sinh\r.
\end{align}
At large $\t$, the saddle point locates at
\begin{align}
\frac{\p \zeta}{\p \hat p^i}=-\frac{\hat p_i\cosh\r}{\sqrt{1+\hat p_i^2}}+n_i\sinh\r=0 \qlq  \hat{p}^\m=-t^\m=(\cosh\r,n^i\sinh\r).
\end{align}
Computing the determinant of the second derivative of the phase $\zeta$
\begin{align}
\det m\frac{\p}{\p \hat p^i}\frac{\p}{\p \hat p^j}\zeta\bigg|_{\hat p^i=-t^i}
&=\det m\bigl(n_in_j\tanh^2\r-\delta_{ij}\bigr)=-\frac{m^3}{\cosh^2\r}
\end{align}
at the saddle point, we eventually arrive at
\begin{align}
\Phi&=\frac{1}{2(2\pi)^{3/2}}[\frac{\sqrt{m}}{\t^{3/2}}a(y)+O(\t^{-5/2})]\e^{-\i m\t}+\hc.
\end{align}
For simplicity, we define
\begin{align}
\varphi(y)=\frac{1}{2(2\pi)^{3/2}}a(y) \qaq \varphi^\dagger(y)=\frac{1}{2(2\pi)^{3/2}}a^\dagger(y)\label{vpexp}
\end{align}
as the leading boundary fields (called also fundamental fields) such that we have \cite{Campiglia:2015kxa}
\begin{align}
\Phi=[\frac{\sqrt{m}}{\t^{3/2}}\vp(y)+O(\t^{-5/2})]\e^{-\i m\t}+\hc.\label{fallPhi}
\end{align}

With the asymptotic form for $\Phi$ and the Laplacian \eqref{lpls}, we can expand the equation of motion. It turns out the leading $\t^{-3/2}$ order is trivial
\begin{align}
(m^2-m^2)\varphi(y)\e^{-\i m\t}+\hc=0,
\end{align}
which implies the fundamental fields are free. The subleading equations of motion can be used to determine subleading fields from the fundamental fields. Note that the interaction terms do not affect the leading-order equations of motion.

\paragraph{Remark.} The fall-off conditions \eqref{fallPhi} can also be obtained by solving the Klein-Gordon equation near $i^+$. We assume the fall-off behaviour as 
\begin{align}
\Phi\sim \tau^{\alpha} \e^{-\i m'\tau}\varphi(y)
\end{align} 
which solves the KG equation order by order. At the leading order, we find 
\begin{align}
m'^2=m^2\quad\Rightarrow\quad m'=\pm m.
\end{align} 
The two branches correspond to the positive and negative frequency modes, respectively. At the subleading order, the constant should be chosen as $\alpha=-\frac{3}{2}$, otherwise one finds $\varphi=0$.

\subsubsection{Stress tensor and Poincar\'e charges}
For the massive scalar, one has the stress tensor below
\begin{align}
T_\mn=\p_\m\Phi\p_\n\Phi-\frac{1}{2}\eta_\mn(\p_\rho\Phi\p^\rho\Phi+m^2\Phi^2)-\eta_{\mu\nu}V[\Phi].
\end{align}
Near $i^+$, the  components relevant to the Poincar\'e charges are  $T^\t{}_\a$ 
\begin{subequations}
\begin{align}
T^\t{}_\t&=-\frac{m^3}{\t^3}2\vp \vp^\dagger+o(\t^{-3}),\\
T^\t{}_a&=\frac{m^2}{\t^3}\big(\i\vp \nabla_a\vp\e^{-2\i m\t}+\i\vp \nabla_a\vp^\dagger\big)+\hc+\cdots.
\end{align}
\end{subequations}
Interestingly, the interaction terms in the potential $V[\Phi]$ do not contribute to the leading order of the components $T^\tau_{\ \tau}$ and $T^\tau_{\ a}$.

Now it is the time to compute the Poincar\'e charges. For $\xi_c=c^\m\p_\m$ with $c^\mu$ a constant vector, the momentum is
\begin{align}
Q_c=\int_{i^+}(\d^3x)_\t\,T^{\t}{}_\a\xi_c^\a=2m^3c^\m\int\d^3y \sqrt{h}\,t_\m\vp^\dagger\vp,\label{Qc}
\end{align}
while for $\xi_\omega=\omega^\mn(x_\m\p_\n-x_\n\p_\m)$ with $\omega^\mn $ a constant skew-symmetric tensor, we have
\begin{align}
Q_\omega&=-m^2\omega^\mn\int\d^3y \sqrt{h}\, X^a_\mn\big(\i\vp \nabla_a\vp\e^{-2\i m\t}+\i\vp \nabla_a\vp^\dagger+\hc\big)
\end{align}
where $X^a_\mn$ is the Killing vector defined in \eqref{Xmna}. It seems that the first term is not conserved due to the oscillating factor. However, it is actually a boundary term since we have $\nabla_aX^a_\mn=0$. To remove the boundary term, the field $\varphi$ should decay as $o(\e^{-\rho})$ near $\rho\to\infty$. Thus, the real Lorentz charge is
\begin{align}
Q_\omega&=-\i m^2\omega^\mn\int\d^3y \sqrt{h}\, X^a_\mn\big(\vp \nabla_a\vp^\dagger-\vp^\dagger \nabla_a\vp\big).
\label{Qomega}
\end{align}

We can rewrite the momentum as
\begin{align}
Q_c=c^\m\int \frac{\d^3p}{(2\pi^3)}\,\frac{m}{2\omega_{\bm p}}t_\m a^\dagger(\bm p)a(\bm p),
\end{align}
where we have used \eqref{vpexp} and
\begin{align}
\d^3p=|\det \p_ap^i|\d^3y=m^3\cosh\r\sqrt{h}\d^3y.
\end{align}
When $c^\m$ takes $(1,\bm 0)$, the energy is recovered
\begin{align}
E=\frac{1}{2}\int \frac{\d^3p}{(2\pi^3)}\, a^\dagger(\bm p)a(\bm p)=\int \frac{\d^3p}{(2\pi^3)}\frac{1}{2\omega_{\bm p}}\, \omega_{\bm p}a^\dagger(\bm p)a(\bm p).
\end{align}

\subsubsection{Poincar\'e charge algebra}
In this subsection, we calculate the commutators between the above charges. To do so, we need the fundamental commutation relation. Taking the variation of the Lagrangian \eq{scalarL} leads to
\begin{align}
\delta\cl=(\Box-m^2)\Phi\delta \Phi-\p_\m(\p^\m\Phi\delta\Phi),
\end{align}
from which we can compute the boundary symplectic form
\begin{align}
\Omega&=-\int_{i^+} (\d^3x)_\t \p^\t\delta \Phi\wedge\delta\Phi\nn\\
&=-\i m^2\int \d^3y \sqrt{h}\,(\delta\vp^\dagger\wedge\delta\vp-\delta\vp\wedge\delta\vp^{\dagger}).
\end{align}
This symplectic form gives rise to the fundamental commutator
\begin{align}
[\vp(y),\vp^\dagger(y')]
&=\frac{1}{2m^2}\delta^{(3)}(y-y'),\label{fundcomscalar}
\end{align}
where the Dirac delta function on $H^3$ satisfies
\begin{align}
\int\d^3y \sqrt{h}\, \delta^{(3)}(y-y')=\int\d^3y\,\delta(\r-\r'){\delta(\th-\th')\delta(\phi-\phi')}=1.
\end{align}
\eq{fundcomscalar} can also be derived from \eq{vpexp} and the canonical commutator
\begin{align}
[a(\bm p),a^\dagger(\bm p')]=(2\pi)^32\omega_{\bm p}\delta(\bm p-\bm p').
\end{align}
With the parameterization $p^i=mn^i\sinh\r$, we find
\begin{align}
\delta(\bm p-\bm p')=|\det \p_ap^i|^{-1}\delta(\r-\r'){ \delta(\th-\th')\delta(\phi-\phi')}=\frac{1}{m^3\cosh\r}\delta^{(3)}(y-y'),
\end{align}
which leads to the commutation relation \eq{fundcomscalar}.

One can check that the commutator between Poincar\'e charges and the fundamental field will produce the boundary Poincar\'e transformation. We first consider
\begin{align}
[Q_c,\vp]= -mc^\m t_\m\vp.
\end{align}
It agrees with the classical variation 
\begin{align}
\delta_c\vp=-\i mc^\m t_\m\vp \label{deltacvp}
\end{align}
which comes from the bulk Lie derivative
\begin{align}
\cl_{\xi_c}\Phi=\frac{m^{1/2}}{\t^{1/2}}(-\i mc^\m t_\m\vp\e^{-\i m\t}+\hc)+\cdots.
\end{align} 
The variation \eqref{deltacvp} is a local rotation with a specified coordinate dependence. For the Lorentz transformation, we have
\begin{align}
[Q_\omega,\vp]
&=-\i\omega^\mn X^a_\mn \nabla_a\vp,\label{qomegacaS}
\end{align}
where the Killing equation of $X^\mn_a$ has been used. One can also compute
\begin{align}
\cl_{\xi_\omega}\Phi=\frac{m^{1/2}}{\t^{1/2}}\omega^\mn X_\mn^a\nabla_a\vp\e^{-\i m\t}+\hc+\cdots,
\end{align}
and hence
\begin{align}
\delta_\omega\vp=\omega^\mn X_\mn^a\nabla_a\vp
\end{align}
which is consistent with \eq{qomegacaS}. Note that $\delta_\omega\vp$ is also the Lie derivative of $\vp$ along $\xi_{\omega}$ on $H^3$. {Note also that the actions of $\i Q_{c/\omega}$ on $\vp$ and $\vp^\dagger$ are conjugate to each other.

We are prepared to compute the charge algebra
\begin{subequations}
\begin{align}
[Q_{c_1},Q_{c_2}]&=0,\\
[Q_c,Q_\omega]&=\i Q_{\tilde{c}},\\
[Q_{\omega_1},Q_{\omega_2}]&=\i Q_{\omega_{12}},
\end{align}
\end{subequations}
with the parameters satisfying
\begin{subequations}
\begin{align}
\tilde{c}^\m t_\m&=-\omega^\mn c^\r X_\mn^a\p_at_\r=\omega^\mn c^\r(h_{\m\r}t_\n-h_{\n\r}t_\m),\label{tildec}\\
\omega_{12}^\mn X_\mn^a&=\omega_1^\mn\omega_2^\rs[X_\mn,X_{\rs}]^a.
\end{align}
\end{subequations}
Here, $[X_\mn,X_{\rs}]$ produces the Lorentz algebra and $h_\mn$ is the bulk version of $h_{ab}$
\begin{align}
h_\mn=h_{ab}S_\m^a S_\n^b=s_\m s_\n+\g_\mn=\eta_{\mu\nu}+t_\mu t_\nu,\label{hmn}
\end{align}
like its partner $\g_\mn$ for $\g_{AB}$ on the unit sphere. The induced metric $h_\mn$ differs from the metric $\eta_\mn$ only by $-t_\m t_\n$, and by symmetry, the latter factor can be added to the right-hand side of \eqref{tildec}. Thus, we obtain what we want
\begin{align}
\tilde{c}^\m t_\m&=\omega^\mn c^\r(\eta_{\m\r}t_\n-\eta_{\n\r}t_\m).
\end{align}
}
In conclusion, the Poincar\'e algebra is indeed realized by our charges. Since the charges only depend on the leading order of the bulk field, the interaction terms do not deform the charge algebra. Therefore, we will only consider free massive fields from now on. 

\subsection{Proca field}
In this subsection, we will extend the previous discussion to the Proca field that describes massive spin 1 field \cite{Proca:1936fbw}. 
\subsubsection{Proca field and its asymptotic expansion}
The Lagrangian for a Proca field with mass $m$ reads
\begin{align}
\cl=-\frac{1}{4}F_\mn F^\mn-\frac{1}{2}m^2A_\m A^\m,\label{procaL}
\end{align}
which leads to the following equation of motion
\begin{align}
\nabla_\m F^\mn-m^2A^\n=0.\label{pmfmn}
\end{align}
Note that the Proca field does not have gauge symmetry and acting \(\nabla^\n\) on \eq{pmfmn}, we find
\begin{align}
\nabla_\m A^\m=0\label{Lorenzgauge}
\end{align}
which takes the same form as the Lorenz gauge. Given this, the equation of motion becomes the Klein-Gordon equation
\begin{align}
(\Box-m^2)A_\m=0.
\end{align}

Near $i^+$, we impose the following large $\t$ expansion
\begin{align}
A_\m=[\frac{\sqrt{m}}{\t^{3/2}}\ca_\m(y)+O(\t^{-5/2})]\e^{-\i m\t}+\hc,\label{Amcam}
\end{align}
which can be derived from the bulk mode expansion
\begin{align}
A_\m=\sum_{\l}\int \frac{\d^3p}{(2\pi)^3}\frac{1}{2\omega_{\bm p}}[\epsilon_\m^\l a_{\l}(\bm p)\e^{\i p\cdot x}+\hc]
\end{align}
by virtue of the saddle-point approximation. For simplicity, we have defined  
\begin{align}
\ca_\m=\frac{1}{2(2\pi)^{3/2}}\sum_{\l}\epsilon_\m^\l(y) a_\l\label{camexp}
\end{align}
in \eqref{Amcam}. Note that the polarization vector $\epsilon_\mu^\lambda$ has three independent modes since it satisfies the Lorenz condition
\begin{align}
p^\mu \epsilon_\mu^{\lambda}(\bm p)=0.
\end{align}
The orthogonality and completeness conditions of the polarization vectors are 
\begin{subequations}\label{orthcomps1}
\begin{align}
\epsilon_\mu^\lambda(\bm p)\epsilon^{*\mu\lambda'}(\bm p)&=\delta^{\lambda\lambda'},\\
\sum_\l\epsilon_\mu^{\lambda}(\bm p)\epsilon_{\nu\lambda}(\bm p')&=\eta_{\mu\nu}+\frac{p_\mu p_\nu}{m^2}=h_{\mu\nu}.
\end{align}
\end{subequations}
Since the four-momenta $p^\mu$ are fixed to be proportional to the normal vector $t^\mu$ in the saddle point approximation 
\begin{align}
p^\mu=-m t^\mu,\label{pmt}
\end{align} 
we can identify $\eta_{\mu\nu}+t_\mu t_\nu$ with the induced metric $h_\mn$ of the hyperboloid when embedded in the Minkowski spacetime, also seeing \eqref{hmn}. 

In the coordinates system of $\{x^\a\}$, we have
\begin{subequations}
\begin{align}
A_\t&=-t^\m A_\m=[\frac{\sqrt{m}}{\t^{3/2}}\ca_\t+O(\t^{-5/2})]\e^{-\i m\t}+\hc \qwq \ca_\t=-t^\m\ca_\m\\
A_{a}&=\t S^\m_a A_\m=[\frac{\sqrt{m}}{\t^{1/2}}\ca_{a}+O(\t^{-3/2})]\e^{-\i m\t}+\hc \qwq \ca_a=S_a^\m \ca_\m.
\end{align}\label{AtAa}
\end{subequations}
In reverse, it gives
\begin{align}
\ca_\m=t_\m\ca_\t+S_\m^a\ca_a.
\end{align}

\subsubsection{Asymptotic equation of motion}
At first, we compute the independent nonvanishing components of the strength tensor. $F_{\t a}$ reads
\begin{align}
  F_{\t a}=-\frac{m^{3/2}}{\t^{1/2}}\i\ca_a\e^{-\i m\t}-\frac{m^{1/2}}{\t^{3/2}}\Bigl(\frac{1}{2}\ca_a +\p_a\ca_\t+\i m\ca_a^{(1)}\Bigr)\e^{-\i m\t}+\hc+\cdots,
\end{align}
where the subleading Proca field is labeled by a superscript $(1)$
\begin{align}
A_\m=\frac{\sqrt{m}}{\t^{3/2}}[\ca_\m(y)+\frac{1}{\t}\ca^{(1)}_\m(y)+O(\t^{-2})]\e^{-\i m\t}+\hc. 
\end{align}
$F_{ab}$ takes the form
\begin{align}
  F_{ab}=\frac{m^{1/2}}{\t^{1/2}}(\nabla_a\ca_b-\nabla_b\ca_a)\e^{-\i m\t}+\hc+\cdots.
\end{align}

Now we expand the equation of motion \eq{pmfmn}. The $\t$ component gives rise to
\begin{align}
0&=[\frac{m^{5/2}}{\t^{3/2}}\ca_\t+\frac{m^{3/2}}{\t^{5/2}}(m\ca_\t^{(1)}-\i \nabla_a\ca^a) +O(\t^{-7/2})]\e^{-\i m\t}+\hc,
\end{align}
while the $a$ component results in
\begin{align}
0&=\p_\t F^{\t a}+\frac{3}{\t}F^{\t a}+\nabla_bF^{ba} -m^2A^a\nn\\
&=[\frac{m^{5/2}}{\t^{5/2}}(\ca^a-\ca^a)-\frac{m^{3/2}}{\t^{7/2}}\i\p^a\ca_\t+O(\t^{-9/2})]\e^{-\i m\t}+\hc.
\end{align}
From the above, one can obtain
\begin{align}
\ca_\t=0 \qaq \ca_\t^{(1)}=\frac{\i}{m}\nabla_a\ca^a.
\label{solution}
\end{align}
These results agree with  \eqref{Lorenzgauge}
\begin{align}
&0=\nabla_\a A^\a=\p_\t A^\t+\nabla_aA^a+\Gamma^{a}_{a\t}A^\t=\p_\t A^\t+\nabla_aA^a+\frac{3}{\t}A^\t\nn\\ 
\Rightarrow\quad& 0=\frac{m^{3/2}}{\t^{3/2}}\i \ca_\t\e^{-\i m\t}+\frac{\sqrt{m}}{\t^{5/2}}(\i m\ca_\t^{(1)}-3\ca_\t+\nabla_a\ca^a)\e^{-\i m\t}+\hc+\cdots.
\end{align}

From the leading order solution of the equation of motion, we conclude that $\mathcal A_a$ is the fundamental field that encodes the three propagating degrees of freedom.

\subsubsection{Stress tensor and Poincar\'e charges}
Given the Lagrangian, the stress tensor of a matter field can be derived from taking the variation with respect to the dynamical metric and then setting the metric back to the background metric. For the Proca field, it takes the form
\begin{align}
T^\mn
&=F^{\m\r}F^\n{}_\r-\frac{1}{4}\eta^\mn (F_\rs)^2+m^2(A^\m A^\n-\frac{1}{2}A_\r A^\r \eta^\mn).\label{mATmn}
\end{align}
With the fall-off \eqref{AtAa}, the relevant components on-shell are
\begin{align}
T^{\t}{}_\t&=-\frac{m^3}{\t^3}2\ca_a\ca^{\dagger a}+O(\t^{-4})
\end{align}
 and
\begin{align}
T^{\t}{}_a&=\frac{m^2}{\t^3}\i\Big[\ca^b(\nabla_a\ca_b-\nabla_b\ca_a) -\ca_a\nabla_b\ca^b\Big]\e^{-2\i m\t}\nn\\
&\quad -\frac{m^2}{\t^3}\i\Big[\ca^{\dagger b}(\nabla_a\ca_b-\nabla_b\ca_a) +\ca^\dagger _a\nabla_b\ca^{b}\Big]+\hc+\cdots.
\end{align}

Now we are prepared to calculate the Poincar\'e charges for the Proca field. For the translation generator $\xi_c=c^\m\p_\m$, we get
\begin{align}
Q_c=\int_{i^+}(\d^3x)_\t\,T^{\t}{}_\a\xi_c^\a=2m^3c^\m\int\d^3y \sqrt{h}\,t_\m\ca_a\ca^{\dagger a}.\label{Qca}
\end{align}
For the Lorentz generator $\xi_\omega=\omega^\mn(x_\m\p_\n-x_\n\p_\m)$, we obtain
\begin{align}
Q_\omega&={-}m^2\omega_\mn\int\d^3y \sqrt{h}\, X_a^\mn\Big[\i Q^{abcd}\ca_b\nabla_c\ca_d\e^{-2\i m\t}-\i P^{abcd}\ca^\dagger_b\nabla_c\ca_d+\hc\Big],
\end{align}
where $X^a_\mn$ was defined in \eqref{Xmna} and we have defined two tensors
\begin{subequations}
\begin{align}
  P_{abcd}&=h_{ab}h_{cd}+h_{ac}h_{bd}-h_{ad}h_{bc},\\ 
  Q_{abcd}&=h_{ac}h_{bd}-h_{ab}h_{cd}-h_{ad}h_{bc}.
\end{align}
\end{subequations}

The first term with the factor $\e^{-2\i m\t}$ can be eliminated. Note that performing integration by parts and discarding the boundary terms lead to
\begin{align}
&\int\d^3y \sqrt{h}\, Q^{abcd}X_a\ca_b\nabla_c\ca_d=\int\d^3y \sqrt{h}\,\big[\ca^a\ca^b\nabla_aX_b-\frac{1}{2}\ca_b\ca^b\nabla_aX^a\big],
\end{align}
which vanishes for the Killing vector $X^a_\mn$. Therefore, the Lorentz charge should be defined as
\begin{align}
Q_\omega&=\i m^2\omega_\mn\int\d^3y \sqrt{h}\, P^{abcd}X^\mn_a\big(\ca^\dagger_b\nabla_c\ca_d-\ca_b\nabla_c\ca^\dagger_d\big).\label{Qom}
\end{align}

\subsubsection{Poincar\'e charge algebra}
Taking the variation of the Lagrangian \eq{procaL} leads to
\begin{align}
\delta\cl=\p_\m F^\mn\delta A_\n-m^2A^\m\delta A_\m-\p_\m(F^\mn\delta A_\n),
\end{align}
from which we can compute the boundary symplectic form
\begin{align}
\Omega&=-\int_{i^+} (\d^3x)_\t \delta F^{\t\a}\wedge\delta A_\a\nn\\
&=-\i m^2\int \d^3y \sqrt{h}\,(\delta\ca_a^\dagger\wedge\delta\ca^a-\delta\ca_a\wedge\delta\ca^{\dagger a}).
\end{align}
It gives rise to the fundamental commutator
\begin{align}
[\ca_a(y),\ca_b^\dagger(y')]
&=\frac{h_{ab}}{2m^2}\delta^{(3)}(y-y').\label{fundcom}
\end{align}
Note that \eq{fundcom} can also be derived from \eq{camexp} and the canonical commutator
\begin{align}
[a_\l(\bm p),a^\dagger_{\l'}(\bm p')]=(2\pi)^32\omega_{\bm p}\delta(\bm p-\bm p')\delta_{\l\l'}.
\end{align}

We first consider
\begin{align}
[Q_c,\ca_a]=- mc^\m t_\m\ca_a,
\end{align}
which agrees with
\begin{align}
\cl_{\xi_c}A_a=\frac{m^{1/2}}{\t^{1/2}}(-\i mc^\m t_\m\ca_a\e^{-\i m\t}+\hc)+\cdots \quad\Rightarrow\quad \delta_c\ca_a=-\i m c^\m t_\m \ca_a.
\end{align}
For the Lorentz transformation, we have
\begin{align}
[Q_\omega,\ca_e]&=-\frac{\i}{2}\omega_\mn [X^\mn_a(P^a{}_e{}^{cd} \nabla_c\ca_d+P^{abc}{}_e\nabla_c\cA_b)+P^{abc}{}_e\nabla_cX^\mn_a\cA_b]\nn\\
&=-\i\omega^\mn(X^a_\mn \nabla_a\ca_e+\ca_a\nabla_eX_\mn^a),\label{qomegaca}
\end{align}
where the Killing equation of $X^\mn_a$ has been used. One can also compute
\begin{align}
&\cl_{\xi_\omega}A_a=\frac{m^{1/2}}{\t^{1/2}}\omega^\mn(X_\mn^b\nabla_b\ca_a+\ca_b\nabla_aX_\mn^b)\e^{-\i m\t}+\hc+\cdots\\
&\Rightarrow\qquad \delta_\omega\ca_a=\omega^\mn(X_\mn^b\nabla_b\ca_a+\ca_b\nabla_aX_\mn^b)
\end{align}
which is consistent with \eq{qomegaca}.

At last, we can compute the algebra formed by these charges. The result is exactly the previous Poincar\'e algebra.

\subsection{Massive Fierz-Pauli field}
\subsubsection{Foundations}
The massive Fierz-Pauli Lagrangian reads \cite{Fierz:1939ix}
\begin{align}
\cl&=-\frac{1}{2}(\partial_\rho H_{\mu \nu} \partial^\rho H^{\mu \nu}- 2 \partial_\mu H^{{\rho\sigma}}\partial_\rho H^{\mu}_{\sigma} + 2\partial_\mu H \partial_\nu H^{\mu \nu} - \partial^\mu H \partial_\mu H)\nn\\
&\quad\, -\frac{1}{2}m^2(H_\mn^2-H^2),
\end{align}
where $H=H_\mn \eta^\mn$. We use $H_\mn$ to represent a symmetric massive spin 2 field since $h_{ab}$ has been used to denote the metric of the unit hyperboloid. Compared to the linearized gravity, we take \(32\pi G=1\) where $G$ is the Newton constant. Note that in order to avoid ghosts \cite{VanNieuwenhuizen:1973fi}, the mass term cannot be deformed to another form, such as $m_1^2 H_{\mu\nu}^2+m_2^2 H^2$. Paying attention to linear theory needs some explanation. It is known that two issues arise in the linearized Fierz-Pauli theory. Firstly, the Fierz-Pauli theory suffers vDVZ discontinuity \cite{vanDam:1970vg,Zakharov:1970cc} since it makes predictions different from those of linearized general relativity even in the limit as the graviton mass goes to zero. This is solved by the Vainshtein mechanism \cite{Vainshtein:1972sx}, which shows that non-linear effects cure the discontinuity. At distances that are below the Vainshtein radius, the non-linear parts dominate and the predictions of the linear theory cannot be trusted. However, in our case, we will focus on the region near $i^+$ which corresponds to $\tau\to\infty$.  In terms of distance $r$, we find $r=\tau\sinh\rho\to\infty$ for any $\rho>0$.  Therefore, it is safe to trust the linear theory in the asymptotic expansion. The second problem is that most of the non-linear extension of Fierz-Pauli theory suffers the ghost problem \cite{Boulware:1972yco}. This has been solved by dRGT theory \cite{deRham:2010ik,deRham:2010kj}. One can find more details on the massive spin-2 field in the reviews \cite{Hinterbichler:2011tt,deRham:2014zqa}. In our case, the non-linear terms do affect the sub-leading fields. However, the Poincar\'e charges still only depend on the leading order fields. 

The equation of motion can be derived from the Lagrangian
\begin{align}
E_\mn\equiv(\Box-m^2)H_\mn-2\p_\r\p_{(\m}H^\r_{\n)}+\p_\m\p_\n H-\eta_\mn[(\Box-m^2) H-\p_\r\p_\s H^\rs]=0.\label{eoms2}
\end{align}
On the other hand, taking the variation of the Lagrangian results in
\begin{align}
\delta\cl=E^\mn\delta H_\mn-\p_\r[(\p^\r H^\mn-2\p^\m H^{\n\r}+\eta^\mn\p_\s H^{\rs}+\eta^{\n\r}\p^\m H-\eta^\mn\p^\r H)\delta H_\mn],
\end{align}
which gives rise to the symplectic form 
\begin{align}
\bm\Omega=-\int (\d^3x)_\t(\p^\t\delta H^\mn-2\p^\m\delta H^{\n\t}+\eta^\mn\p_\s \delta H^{\t\s}+\eta^{\n\t}\p^\m\delta H-\eta^\mn\p^\t\delta H)\wedge\delta H_\mn,
\end{align}
on a constant $\t$ surface.

\paragraph{Simplifications.}
If acting on \eq{eoms2} with $\p^\m$ and assuming $m^2\ne 0$ \cite{Hinterbichler:2011tt}, we will get
\begin{align}
\p^\m H_\mn-\p_\n H=0,\label{gauges2}
\end{align}
which is similar to the de Donder gauge. Plugging \eqref{gauges2} into \eqref{eoms2}, we  find
\begin{align}
\Box H_\mn-\p_\m\p_\n H-m^2(H_\mn-\eta_\mn H)=0,
\end{align}
whose trace leads to $H=0$. Given \eq{gauges2}, we arrive at two important simplifying conditions
\begin{align}
H=0 \qaq \p^\m H_\mn=0,\label{TT}
\end{align}
which can be interpreted as the transverse and traceless conditions. They can simplify the EOM to the Klein-Gordon equation
\begin{align}
(\Box-m^2)H_\mn= 0.\label{KG}
\end{align}
Such $H_\mn$ has 5 degrees of freedom and describes the massive spin-2 irreducible representation of the Poincar\'e group. From now on, we use the notation $\approx$ to indicate that the equation is valid on-shell, namely under \eqref{TT} and \eqref{KG}.

\paragraph{Stress tensor I.}
Variating the action with respect to the metric gives rise to the stress tensor. To avoid overlooking terms, we write the (covariant) Fierz-Pauli action as
\begin{align}
S=-\frac{1}{2}\int \d^4 x\sqrt{-g}[L^{\mu_1\cdots\mu_6}\nabla_{\mu_1}H_{\mu_2\mu_3}\nabla_{\mu_4}H_{\mu_5\mu_6}+m^2(H_{\mu\nu}H^{\mu\nu}-H^2)]
\end{align}  
where the tensor $L^{\mu_1\cdots\mu_6}$ is defined as
\begin{align}
L^{\mu_1\cdots\mu_6}=&\,\frac{1}{2}(g^{\mu_1\mu_4}g^{\mu_2\mu_5}g^{\mu_3\mu_6}+g^{\mu_1\mu_4}g^{\mu_2\mu_6}g^{\mu_3\mu_5})\nn\\
&-\frac{1}{2}(g^{\mu_1\mu_5}g^{\mu_2\mu_4}g^{\mu_3\mu_6}+g^{\mu_1\mu_5}g^{\mu_3\mu_4}g^{\mu_2\mu_6}+g^{\mu_1\mu_6}g^{\mu_2\mu_4}g^{\mu_3\mu_5}+g^{\mu_1\mu_6}g^{\mu_3\mu_4}g^{\mu_2\mu_5})\nn\\
&+(g^{\mu_1\mu_5}g^{\mu_2\mu_3}g^{\mu_4\mu_6}+g^{\mu_1\mu_6}g^{\mu_2\mu_3}g^{\mu_4\mu_5})-g^{\mu_1\mu_4}g^{\mu_2\mu_3}g^{\mu_5\mu_6}.
\end{align} 
This tensor is symmetric under $\mu_2\leftrightarrow\m_3$ and $\mu_5\leftrightarrow\m_5$, as well as $\m_1\mu_2\m_3\leftrightarrow\m_4\m_5\m_6$.  Variating the action with respect to the metric gives rise to the stress tensor
\begin{align}
T_{\mu\nu}&=-\frac{2}{\sqrt{-g}}\frac{\delta S}{\delta g^\mn}\Big|_{g\to\eta}\nn\\
&=T_{\mu\nu}^{(1)}+T_{\mu\nu}^{(2)}+T_{\mu\nu}^{(3)}+T_{\mu\nu}^{(4)},
\end{align} 
where the results are divided into 4 parts, coming from the variation of (1) the tensor $L^{\mu_1\mu_2\cdots\mu_6}$, (2) the mass term, (3) the factor $\sqrt{-g}$, and (4) the Christoffel symbols, respectively. 

The evaluation of the first three terms is easy to do and leads to
\begin{align}
T_\mn^{(1+2+3)}&\approx-\frac{1}{2}\eta_\mn(\p_\k H_{\rho\sigma}\p^\k H^{\rho\sigma}-2 \p_{\rho}H_{\sigma\k}\p^\sigma H^{\rho\k})+\p_\mu H_{\rho\sigma}\p_\nu H^{\rho\sigma}+2\p_\rho H_\mu^{\sigma}\p^\rho H_{\n\s}\nn\\
&\quad -2(\p_\mu H^{\rho\sigma}\p_\rho H_{\n\sigma}+\p_\nu H^{\rho\sigma}\p_\rho H_{\m\sigma}+\p_\rho H_{\m\sigma}\p^\sigma H^{\rho}_\n)\nn\\
&\quad -\frac{1}{2}m^2\eta_\mn H_\rs H^\rs+2m^2H_{\m\r} H_\n^{\r}\label{2.111}
\end{align}
under the conditions \eq{TT}.
As for the last term, we obtain
\begin{align}
T_\mn^{(4)}&=\Big[-\nabla_{\s} (g_{\r(\n}\frac{\p\cl}{\p\Gamma^{\m)}_\rs})-\nabla_{\r} (g_{\s(\n}\frac{\p\cl}{\p\Gamma^{\m)}_\rs})+\nabla^\k(g_{\r(\m}g_{\n)\s}\frac{\p\cl}{\p\Gamma^\k_\rs})\Big]_{g\to\eta}\nn\\
&=-2\p_\s[(L_{(\m}{}^{\s\m_3\cdots\m_6}H_{\n)\m_3}+L^\s{}_{(\m}{}^{\m_3\cdots\m_6}H_{\n)\m_3})\p_{\m_4}H_{\m_5\m_6}]\nn\\
&\quad +2\p_\s(L_{(\mn)}{}^{\m_3\cdots\m_6}H^\s_{\m_3}\p_{\m_4}H_{\m_5\m_6})\nn\\
&\approx-2H^{\rs}\p_\r\p_\s H_{\mn} +4\p_\r H_\m^{\s}\p_\s H^{\r}_\n,
\end{align}
where we have used the variation of the Christoffel symbol 
\begin{align}
\delta\Gamma^\mu_{\rho\sigma}=-\frac{1}{2}(g_{\n\r}\nabla_\s\delta g^{\mn}+g_{\n\s}\nabla_\r\delta g^{\mn}-g_{\r\k}g_{\s\l}\nabla^\m\delta g^{\k\l}).
\end{align}
Combining the results (and raising the indices), we obtain the following stress tensor 
\begin{align}
T^\mn &\approx -\frac{1}{2}\eta^\mn(\p_\k H_{\rho\sigma}\p^\k H^{\rho\sigma}+m^2H_\rs^2-2 \p_{\rho}H_{\sigma\k}\p^\sigma H^{\rho\k})+2m^2H^\m_\r H^{\n\r}\nn\\
&\quad +\p^\mu H_{\rho\sigma}\p^\nu H^{\rho\sigma}+2\p_\rho H^{\mu}_\sigma(\p^\rho H^{\n\sigma}+\r\leftrightarrow\s)-2(\p^\mu H^{\rho\sigma}\p_\rho H^\nu_{\sigma}+\m\leftrightarrow\n)\nn\\
&\quad -2H^{\rs}\p_\r\p_\s H^{\mn}.\label{TmnmgTT}
\end{align}
This result agrees with the one in the literature \cite{Petrov:2017bdx}. One can easily verify that it is on-shell conserved
\begin{align}
\p_\m T^\mn\approx0 .
\end{align}

\paragraph{Stress tensor II.}
We can also use Noether's theorem to derive the canonical stress tensor
\begin{align}
\Th^\mn&=-\frac{\p\cl}{\p\p_\m H_\rs}\p^\n H_\rs+\eta^\mn\cl\nn\\
&\approx\p^\m H^\rs\p^\n H_\rs-2\p^{\r}H^{\s\m}\p^\n H_\rs\nn\\
&\quad -\frac{1}{2}\eta^\mn(\p_\k H_{\rho\sigma}\p^\k H^{\rho\sigma}+m^2H_\rs^2-2 \p_{\rho}H_{\sigma\k}\p^\sigma H^{\rho\k}),
\end{align}
which is conserved but not symmetric. The symmetrization can be done through the Belinfante correcting method \cite{Petrov:2017bdx}. The correcting term is constructed from the spin angular momentum current. Therefore, we consider an infinitesimal Lorentz transformation $\L_\m{}^\n=\delta_\m^\n+\delta\omega_\m{}^\n$ under which
\begin{align}
\delta_\omega H_\mn(x)&=H'_\mn(x)-H_\mn(x)\nn\\
&=-\delta\omega_\m{}^\r H_{\r\n}(x)-\delta\omega_\n{}^\r H_{\r\m}(x)-\delta\omega^\r{}_\s x^\s\p_\r H_\mn(x)+O(\delta\omega^2).
\end{align} 
The spin angular momentum current is given by
\begin{align}
S^{\mn\r}=4\p^\m H^{\s[\n}H^{\r]}_\s-2(\p^{\n}H^{\m\s}H^{\r}_\s+\p^\s H^{\mn}H^{\r}_\s-\n\leftrightarrow\r).\label{smnr}
\end{align}
One can verify that
\begin{align}
\Th^\mn-\Th^\nm=-4\p^{\r}H^{\s[\m}\p^{\n]} H_\rs\approx-\p_\r S^{\r\mn}
\end{align}
Thus, the stress tensor can be symmetrized 
\begin{align}
T^\mn=\Theta^\mn+\p_\r B^{\r\mn},
\end{align}
where $B^{\r\mn}$ is the Belinfante tensor
\begin{align}
B^{\r\mn}=\frac{1}{2}(S^{\r\mn}+S^{\mn\r}-S^{\n\r\m}).
\end{align}
We eventually obtain \eqref{TmnmgTT} as expected.

\subsubsection{Asymptotic expansion}
Similar to the massive scalar and vector fields, we impose the fall-off conditions near $i^+$
\begin{align}
H_\mn(x)=[\frac{\sqrt{m}}{\t^{3/2}}\cH_\mn(y)+O(\t^{-5/2})]\e^{-\i m\t}+\hc
\end{align} 
which could be verified by the standard mode expansion
\begin{align}
H_\mn(x)=\sum_\l\int \frac{\d^3p}{(2\pi)^3}\frac{1}{2\omega_{\bm p}}[a_\l(\bm p)\ep^\l_\mn(\bm p)\e^{\i p\cdot x}+\hc]\label{hmnexp}
\end{align} 
and using the saddle point approximation. The polarization tensor satisfies the constraints 
\begin{align}
\epsilon^{\lambda}_{\mu\nu}(\bm p)=\epsilon^{\lambda}_{\nu\mu}(\bm p),\quad p^\mu \epsilon^{\lambda}_{\mu\nu}(\bm p)=0,\quad \eta^{\mu\nu}\epsilon_{\mu\nu}^\lambda(\bm p)=0.
\end{align}
A general representation of the polarization tensors satisfies the orthonormality condition 
\begin{align}
\epsilon_{\mu\nu}^\lambda(\bm p)\epsilon^{*\mu\nu}_{\lambda'}(\bm p)=\delta^\lambda_{\lambda'}
\end{align} 
and the completeness relation 
\begin{align}
\sum_\l\epsilon_{\mu\nu}^\lambda(\bm p)\epsilon^*_{\rho\sigma,\lambda}(\bm p)= \frac{1}{2} (h_{\mu\rho}h_{\nu\sigma}+h_{\mu\sigma}h_{\nu\rho}-\frac{2}{3} h_{\mu\nu}h_{\rho\sigma}).\label{completes2}
\end{align}
One can check that the tensor structure on the right-hand side fulfills the symmetric, transverse, and traceless conditions.

Now one can derive
\begin{align}
\p_\m H_\rs&=[-\i \frac{m^{3/2}}{\t^{3/2}}t_\m \cH_\rs+\frac{\sqrt{m}}{\t^{5/2}}(S^a_\m\p_a\cH_\rs-\frac{3}{2}t_\m \cH_\rs-\i mt_\m \cH_\rs^{(1)})\nn\\
&\quad +O(\t^{-7/2})] \e^{-\i m\t}+\hc,
\end{align}
and thus
\begin{align}
\p_\n\p_\m H_\rs&=[-\frac{m^{5/2}}{\t^{3/2}}t_\m t_\n \cH_\rs-\i\frac{m^{3/2}}{\t^{5/2}}(2t_{(\n} S^a_{\m)}\p_a\cH_\rs-(\eta_\mn+4t_\m t_\n) \cH_\rs\nn\\
&\quad\ -\i mt_\m t_\n\cH_\rs^{(1)})+O(\t^{-7/2})] \e^{-\i m\t}+\hc,
\end{align}
where the superscript $(1)$ labels the subleading field
\begin{align}
H_\mn=\frac{\sqrt{m}}{\t^{3/2}}[\ch_\mn(y)+\frac{1}{\t}\ch^{(1)}_\mn(y)+O(\t^{-2})]\e^{-\i m\t}+\hc. 
\end{align}
It follows that
\begin{align}
\Box H_\rs&=\frac{m^{5/2}}{\t^{3/2}}[\cH_\rs+\frac{1}{\t} \cH_\rs^{(1)}+O(\t^{-2})] \e^{-\i m\t}+\hc,
\end{align}
and the Klein-Gordon equation is satisfied at the leading order without imposing any additional conditions for the leading and subleading fields. On the other hand, the transverse condition gives rise to
\begin{align}
t_\m \cH^\mn=0 \qaq S_\m^a\p_a \cH^\mn-\i mt_\m \cH^{(1)\mn}=0.\label{2.120}
\end{align}
The first condition implies that
\begin{align}
\cH^{\t\n}=\frac{\p\t}{\p x^\m}\cH^\mn=t_\m \cH^\mn=0,\label{transcH}
\end{align}
namely, only $\ch^{ab}$ does not vanish, while the second equation links the subleading field to the leading one. The traceless condition leads to
\begin{align}
\mathcal H^{\mu}_{\ \mu}=0,\qquad \mathcal H^{(1)\mu}_{\ \mu}=0.
\end{align} 
Combining with the condition \eqref{transcH}, we find that $\mathcal H_{ab}$ is also traceless
\begin{align}
h^{ab}\cH_{ab}=0.
\end{align}
Such a $\mathcal H_{ab}$ has 5 degrees of freedom as expected.

To derive the charges, we need the following components of the stress tensor
\begin{align}
T^\t{}_\t=-t_\m t^\n T^\m{}_\n=-\frac{m^3}{\t^3}2\cH_\mn^\dagger\ch^\mn+O(\t^{-4}),
\end{align}
and
\begin{align}
T^\t{}_a&=\t t_\m S_a^\n T^\m{}_\n\nn\\
&=\frac{m^2}{\t^3}[2m t^\r S_a^\n \ch^{\dagger(1)}_\rs\ch^{\s}_\n-\i \cH^{\dagger\mn}\p_a\cH_\mn+2\i S_a^\n S^b_\r\ch^{\dagger\rs}\p_b\cH_{\n\s}\nn\\
&\qquad\quad -(2m t^\r S_a^\n \ch^{(1)}_\rs\ch^{\s}_\n-\i \cH^{\mn}\p_a\cH_\mn-2\i S_a^\n S^b_\r\ch^{\rs}\p_b\cH_{\n\s})\e^{-2\i m\t}\nn\\
&\qquad\quad+\hc]+\cdots.
\end{align} 
Here, we have used the asymptotic expansion of the stress tensor in the Cartesian frame
\begin{align}
T^\mn&= \frac{m^3}{\t^3}(t^\m t^\n \cH_\rs^\dagger\ch^{\rs}+\hc)\nn\\
&\quad +\frac{m^2}{\t^4}\Big[\eta^\mn(\cdots)+2\i t^{(\m}  \ch^\dagger_\rs(S^{\n) a}\p_a\ch^\rs-\i mt^\n\ch^{(1)\rs})\nn\\
&\qquad\qquad-2(\i t^\m S^a_\r\ch^{\dagger\rs}\p_a\ch^\n_\s-\i t_\r S^{\m a}\ch^{\n}_\s\p_a\ch^{\dagger\rs} +mt^\m t_\r \ch^{\dagger(1)\rs} H^\n_\s+\m\leftrightarrow\n)\nn\\
&\qquad\qquad +2\i t_\r S_\s^a(\ch^{\dagger\m\s}\p_a\ch^{\r\n}+\m\leftrightarrow\n)+2\ch^{\dagger(1)\rs}t_\r t_\s \ch^\mn +\hc \Big]+O(\t^{-5})\nn\\
&\quad +\e^{-2\i m\t}(\cdots)+\e^{2\i m\t}(\cdots),
\end{align}
where we have omitted the unimportant or similar terms for brevity.

At last, we compute the leading order of the symplectic form
\begin{align}
\bm\Omega(\delta\ch,\delta\ch)=- \i m^2\int\d^3y \sqrt{h}\,(\delta\ch^{\dagger ab}\wedge\delta\ch_{ab}-\delta\ch^{ab}\wedge\delta\ch^\dagger_{ab}),\label{sympforms2}
\end{align}
where we have used transverse and traceless conditions. \eqref{sympforms2} leads to the following fundamental commutator
\begin{align}
[\ch_{ab}(y),\ch_{cd}^\dagger(y')]&=\frac{1}{4m^2}(h_{ac}h_{bd}+h_{ad}h_{bc}-\frac{2}{3}h_{ab}h_{cd})\delta^{(3)}(y-y').\label{fundcoms2}
\end{align}
Note that both sides are symmetric and traceless with respect to $ab$ and $cd$. The same structure also appears in the completeness relation of the polarization tensor, i.e., \eqref{completes2}. We will define the following tensor
\begin{align}
\mathcal P_{abcd}=\frac{1}{2}(h_{ac}h_{bd}+h_{ad}h_{bc}-\frac{2}{3}h_{ab}h_{cd})
\end{align}
for simplicity, such that
\begin{align}
[\ch_{ab}(y),\ch_{cd}^\dagger(y')]&=\frac{1}{2m^2}\mathcal P_{abcd}\delta^{(3)}(y-y').\label{fundcoms21}
\end{align}

\subsubsection{Charges and commutators}
Now we are prepared to calculate the charges
\begin{align}
Q_c=-\t^3\int\d^3y \sqrt{h}\,T^{\t}{}_\t\, \xi_c^\t=2m^3c^\m\int\d^3y\sqrt{h}\,t_\m\cH^\dagger_{ab}\ch^{ab},
\end{align}
and
\begin{align}
Q_{\omega}&=-\t^3\int\d^3y \sqrt{h}\,T^{\t}{}_a\, \xi_\omega^a\nn\\
&=m^2\omega^\mn\int\d^3y \sqrt{h}\,X_\mn^a\Big[2\i S_a^\m S^b_\r\ch^{\dagger\s}_\m\p_b\cH^\r_{\s}+\i\cH^{\dagger\mn}\p_a\cH_\mn-2\i S_a^\m S^b_\r\ch^{\dagger\rs}\p_b\cH_{\m\s}\nn\\
&\qquad\qquad\qquad -(2\i S_a^\n S^b_\r  \ch^{\s}_\n\p_b\ch^\r_\s+\i \cH^{\mn}\p_a\cH_\mn+2\i S_a^\n S^b_\r\ch^{\rs}\p_b\cH_{\n\s})\e^{-2\i m\t}+\hc\Big]\nn\\
&=m^2\omega^\mn\int\d^3y \sqrt{h}\,X_\mn^a\Big[2\i\ch^{\dagger}_{ab}\nabla_c\ch^{bc}+ \i\cH^{\dagger bc}\nabla_a\cH_{bc}-2\i\ch^{\dagger bc}\nabla_b\cH_{ac}\nn\\
&\qquad\qquad\qquad -(2\i\ch_{ab}\nabla_c\ch^{bc}+ \i\cH^{bc}\nabla_a\cH_{bc}+2\i\ch^{bc}\nabla_b\cH_{ac})\e^{-2\i m\t}+\hc\Big],\label{QXs2}
\end{align}
where we have used \eqref{2.120} and the identity
\begin{align}
\nabla_aS_\m^b=-\delta_{a}^b t_\m.
\end{align}
Note that the terms with factor $\e^{\pm2\i m\t}$ vanish since they can be combined as a total derivative due to the Killing equation.

Using the fundamental commutator \eqref{fundcoms2}, one can compute the commutator between $Q_c$ and $\ch_{ab}$
\begin{align}
[Q_c,\ch_{ab}]=-mc^\m t_\m \ch_{ab},
\end{align}
which agrees with the classical variation
\begin{align}
\delta_c\ch_{ab}=-\i m c^\m t_\m \ch_{ab}
\end{align}
that comes from the Lie derivative. We also find
\begin{align}
[Q_\omega,\ch_{ab}]=-\i\omega^\mn(X_\mn^c\nabla_c\cH_{ab}+\cH_{ac}\nabla_bX_\mn^c+\cH_{cb}\nabla_aX_\mn^c)
\end{align}
which is consistent with
\begin{align}
\cl_{\xi_\omega}H_{ab}&=\frac{m^{1/2}}{\t^{3/2}}\omega^\mn(X^c_\mn\nabla_c\cH_{ab}+\cH_{ac}\nabla_bX_\mn^c+\cH_{cb}\nabla_aX_\mn^c)\e^{-\i m\t}\nn\\
&\quad +\hc+\cdots
\end{align}
and thus 
\begin{align}
\delta_\omega\ch_{ab}=\omega^\mn(X_\mn^c\nabla_c\cH_{ab}+\cH_{ac}\nabla_bX_\mn^c+\cH_{cb}\nabla_aX_\mn^c).
\end{align}

To close this section, we have verified that charges $Q_c$ and $Q_\omega$ indeed realize the Poincar\'e algebra.

\section{Extended charge algebra at future timelike infinity}\label{char}
In this section, we will extend the Poincar\'e algebra at $i^+$.
\subsection{Massive scalar}
We extract the energy and angular momentum density of the scalar field at $i^+$ from \eqref{Qc} and \eqref{Qomega} correspondingly
\begin{subequations}
\begin{align}
T(y)&=2m^3\varphi^\dagger \varphi,\\ 
M_a(y)&=-\i m^2(\nabla_a\varphi^\dagger \varphi-\varphi^\dagger\nabla_a\varphi).
\end{align}\label{density}
\end{subequations}
Physically, they are well-defined composite operators at $i^+$. Noting that the fundamental field $\varphi$ is proportional to the annihilation operator at $i^+$, we may define the vacuum state at $i^+$ as 
\begin{align}
\varphi|0\rangle=0.
\end{align} 
The vacuum state $|0\rangle$ should be distinguished from the true vacuum state $|\bm 0\rangle$ at a finite time whose form depends on the interactions. However, it is fine to focus on the state $|0\rangle$ in this work. The excited states are obtained by acting on the operator $\varphi^\dagger$ recursively
\begin{align}
\prod_{I}[\varphi^{\dagger}(y_I)]^{m_I}|0\rangle
\end{align} 
where $m_I$ is a non-negative integer whose subscript $I$ denotes the operator inserted at $y_I$. The two-point function follows from the commutator \eqref{fundcomscalar}
\begin{subequations}
\begin{align}
&\langle0|\varphi(y)\varphi(y')|0\rangle=\langle 0|\varphi^\dagger(y)\varphi^\dagger(y')|0\rangle=0,\\ 
&\langle0|\varphi(y)\varphi^\dagger(y')|0\rangle=\frac{1}{2m^2}\delta^{(3)}(y-y').
\end{align}
\end{subequations}

The densities \eqref{density} are lifted to smeared operators on $i^+$ 
\begin{subequations}\label{TfMXs=0}
\begin{align}
\mathcal T_f&=\int \d^3y \sqrt{h}\, f(y):T(y):,\\
\mathcal M_X&=\int \d^3 y\sqrt{h}\, X^a(y):M_a(y):
\end{align} 
\end{subequations}
where $f(y)$ is any smooth function on $H^3$ and $X^a(y)$ is any smooth vector field on $H^3$. The normal order $:\cdots:$ is used to arrange the positions of $\varphi$ and $\varphi^\dagger$ such that the vacuum expectation value of $\mathcal T_f$ and $\mathcal M_X$ vanish. We find the action of the smeared operators on the fundamental field 
\begin{subequations}
\begin{align}
[\mathcal T_f,\varphi(y)]&=-mf(y)\varphi(y),\\ 
[\mathcal M_X,\varphi(y)]&=-\i X^a(y)\nabla_a\varphi(y)-\frac{\i}{2}\varphi(y)\nabla_aX^a(y).\label{MXvp}
\end{align}\label{comTfMXvarphi}
\end{subequations}
Geometrically, we may also uplift the Killing vectors in the bulk to the following form 
\begin{subequations}
\begin{align}
\xi_f&=f\partial_\tau-\frac{1}{\tau}\nabla^af\partial_a, \\ \xi_X&=X^a\partial_a.
\end{align} \label{liftfX}
\end{subequations}
This is motivated by the fact that $\xi_f$ reduces to the translation generator when $f=c^\mu t_\mu$ and $\xi_X$ reduces to the Lorentz transformation generator when $X^a=\omega^{\mu\nu}X_{\mu\nu}^a$. Interestingly, one can also compute the variation of $\varphi$ induced by the bulk vectors 
\begin{subequations}
\begin{align}
&\mathcal L_{\xi_f}\Phi=\xi_f^\a\p_\a \Phi\quad \Rightarrow\quad \delta_f\varphi=-\i mf\varphi,\\
&\mathcal{L}_{\xi_X}\Phi=\xi_X^\a\p_\a \Phi\quad\Rightarrow\quad \delta_X\varphi=X^a\partial_a\varphi.
\end{align} 
\end{subequations}
The commutators \eqref{comTfMXvarphi} match the above variation induced by the Lie derivative, i.e.,
\begin{align}
[\i\mathcal T_f,\cdots]\Leftrightarrow \delta_f(\cdots),\qquad [\i\mathcal M_X,\cdots]\Leftrightarrow\delta_X(\cdots),
\end{align} 
provided that the vector field $X^a$ is divergence-free on $H^3$
\begin{align}
\nabla_aX^a=0. \label{div}
\end{align} 
For this reason, the operators $\mathcal T_f$ and $\mathcal M_X$ may be called charges, even though they are not conserved charges in general.

Note that one can find another vector
\begin{align}
\tilde\xi_X=\frac{\i}{2m}\nabla_aX^a\p_\t+X^a\partial_a\label{tildexiX}
\end{align}
which leads to the variation
\begin{align}
\cl_{\tilde\xi_X}\Phi&=\frac{m^{1/2}}{\t^{3/2}}(X^a\nabla_a\vp+\frac{1}{2}\nabla_aX^a\vp)\e^{-\i m\t}+\hc+\cdots\\
\Rightarrow\tilde{\delta}_X\vp&=X^a\nabla_a\vp+\frac{1}{2}\nabla_aX^a\vp.
\end{align}
This result agrees with \eqref{MXvp} on its original form. The vector \eqref{tildexiX} has a similar structure to the leading order of the superrotation vector
\begin{align}
\wt\xi_Y=\frac{1}{2}\nabla_AY^A(u\p_u-r\p_r)+Y^A\p_A+\cdots
\end{align}
in the sense of the generalized BMS group. However, this vector field is not real due to the factor $\i$. Given that, we prefer the previous $\xi_X$ with $X^a$ to be divergence-free.

Furthermore, it is easy to find that $\tilde\xi_X$ will lead to a charge different from $\cm_X$, but the difference takes the form of 
\begin{align}
\ct_{f=\frac{\i}{2m}\nabla_aX^a}
\end{align} 
which is non-Hermitian. Therefore, we still propose the extended charge operators \eqref{TfMXs=0}. 

Now it is straightforward to obtain the following charge algebra 
\begin{subequations}
\begin{align}
[\mathcal T_{f_1},\mathcal T_{f_2}]&=0,\\
[\mathcal T_f,\mathcal M_X]&=-\i\mathcal T_{X(f)},\label{TfMXscalar}\\
[\mathcal M_X,\mathcal M_Y]&=\i\mathcal M_{[X,Y]}\label{MXMYscalar}
\end{align} 
\end{subequations}
where $f_1,f_2$ are smooth function on $H^3$ and $X,Y$ are smooth divergence-free vectors on $H^3$. Therefore, the group generated by $\mathcal T_f$ may be denoted by $C^{\infty}(H^3)$ and the group generated by $\mathcal M_X$ may be denoted by $\text{MDiff}(H^3)$, which is a shorthand of magnetic diffeomorphism generated by all divergence-free smooth vectors following the terminology of \cite{Guo:2024qzv}. In summary, the charge algebra generated by $\mathcal T_f$ and $\mathcal M_X$ is 
\begin{align}
\text{MDiff}(H^3)\ltimes C^{\infty}(H^3).
\end{align}
Note that there is no central extension for this charge algebra. 

\subsection{Proca field}
The previous discussion can be extended to massive spinning fields. To show this, we still extract the energy and angular momentum density of the Proca field from \eqref{Qca} and \eqref{Qom}
\begin{subequations}
\begin{align}
T(y)&=2m^3\ca_a^\dagger \ca^a,\\ 
M_a(y)&=-\i m^2P^{abcd}\big( \ca_b\nabla_c\ca^\dagger_d- \ca^\dagger_b\nabla_c\ca_d\big),
\end{align} 
\end{subequations}
and then lift the Poincar\'e charges to the following charges 
\begin{subequations}
\begin{align}
\mathcal T_f&=\int\d^3y \sqrt{h}\,f(y):T(y):,\\ 
\mathcal M_X&=\int\d^3y \sqrt{h}\, X^a(y):M_a(y):.\label{MXs=1}
\end{align} \label{tfmxdefs1}
\end{subequations}
The action of the extended charges on the fundamental field is 
\begin{subequations}
\begin{align}
[\i\mathcal T_f,\ca_a]&=-\i mf\ca_a,\\
[\i\mathcal M_X,\ca_a]&=X^b\nabla_b\ca_a+\ca^b\nabla_{[a}X_{b]},
\end{align} \label{comTA}
\end{subequations}
where the divergence-free condition \eqref{div} has been imposed.

Still using \eqref{liftfX}, the first equation of \eqref{comTA} can match the variation induced by the bulk Lie derivative 
\begin{align}
\cl_{\xi_f}A_a=\xi_f^\a\p_\a A_a+\partial_a\xi_f^\a A_\a \quad\Rightarrow\quad \delta_f\ca_a=-\i mf\ca_a
\end{align}
while the latter one fails 
\begin{align}
\cl_{\xi_X}A_a=\xi_X^\a\p_\a A_a+\partial_a\xi_X^\a A_\a \quad \Rightarrow\quad\delta_X\ca_a=X^b\nabla_b\ca_a+\ca_b\nabla_aX^b.
\end{align} 
This mismatch can be cured by introducing the so-called covariant variation 
\begin{align}
\Delta_X \ca_a=\delta_X\ca_a-\frac{1}{2}\delta_X h_{ab} \ca^b=X^b\nabla_b\ca_a+\ca^b\nabla_{[a}X_{b]}=[\i\mathcal M_X,\ca_a],
\end{align} 
where we have used the variation of the boundary metric induced by the bulk Lie derivative
\begin{align}
{ \cl_{\xi_X}\eta_{\ab}=2\nabla_{(\a}X_{\b)} \qrq }\delta_Xh_{ab}=2\nabla_{(a}X_{b)}.
\end{align} 
The advantage of covariant variation is that its action on the boundary field could match the commutator, and it preserves the boundary metric 
\begin{align}
\Delta_X h_{ab}=0.
\end{align}
As an aside, if we do not impose the divergence-free condition, then
\begin{align}
[\i\mathcal M_X,\ca_a]&=X^b\nabla_b\ca_a+\ca^b\nabla_{[a}X_{b]}+\frac{1}{2}\ca_a\nabla_bX^b,
\end{align}
which corresponds to the covariant variation induced by \eqref{tildexiX}. This kind of $\tilde{\xi}_{X}$ has been ruled out by the reality condition. 

Similarly, one can also find
\begin{subequations}
\begin{align}
[\i\mathcal T_f,\ca^\dagger_a]&=\delta_f\ca_a^\dagger=\i mf\ca_a^\dagger,\\
[\i\mathcal M_X,\ca_a^\dagger]&=\Delta_X\ca_a^\dagger=X^b\nabla_b\ca^\dagger_a+\ca^{\dagger b}\nabla_{[a}X_{b]},
\end{align} 
\end{subequations}
Note that we can rewrite the extended charges in terms of their actions on fundamental fields
\begin{subequations}
\begin{align}
\mathcal T_f&=-\i m^2\int\d^3y \sqrt{h}\,(\ca^{a}\delta_f\ca^\dagger_a-\ca^{\dagger a}\delta_f\ca_a),\\ 
\mathcal M_X&=-\i m^2\int\d^3y \sqrt{h}\,(\ca^{a}\Delta_X\ca^\dagger_a-\ca^{\dagger a}\Delta_X\ca_a).
\end{align} \label{rewrite Tf Mx s1}
\end{subequations}

\paragraph{Non-closure.}
We can compute the commutator of the boundary covariant variation
\begin{align}
[\Delta_X,\Delta_Y]\ca_a=\Delta_{[X,Y]}\ca_a-o_{ab}(X,Y)\ca^b,\label{DXDYA1}
\end{align}
where the antisymmetric tensor $o_{ad}$ is defined as
\begin{align}
o_{ad}(X,Y)=[\nabla_{(a}X_{b)}\nabla_{(c}Y_{d)}-\nabla_{(a}Y_{b)}\nabla_{(c}X_{d)}]h^{bc},\label{oabdef}
\end{align}
which vanishes for the Killing vector on $H^3$. Quantum mechanically, \eqref{DXDYA1} corresponds to
\begin{align}
[\cm_X,\cm_Y]=\i \cm_{[X,Y]}- m^2\int \d^3y \sqrt{h}\,o_{ab}(\ca^a\ca^{\dagger b}-\ca^{\dagger a}\ca^b).\label{MXMY1}
\end{align}
We can explicitly compute
\begin{align}
[\cm_X,\cm_Y]&=-\i m^2\int\d^3y \sqrt{h}\,[\cm_X, \ca^a\Delta_Y\ca_a^\dagger-\ca^{\dagger a}\Delta_Y\ca_a]\nn\\ 
&=-m^2\int\d^3y \sqrt{h}\,\big[\Delta_X\ca^a\Delta_Y\ca_a^\dagger+\ca^a\Delta_Y\Delta_X\ca_a^\dagger-\hc\big],
\end{align}
where
\begin{align}  
&\int\d^3y \sqrt{h}\,\big[\Delta_X\ca^a\Delta_Y\ca_a^\dagger-\Delta_X\ca^{\dagger a}\Delta_Y\ca_a\big]\nn\\  &=\int\d^3y \sqrt{h}\,\big[(X^b\nabla_b\ca^a+\nabla^{[a}X^{b]}\ca_b)\Delta_Y\ca^\dagger_a-\hc\big]\nn\\
&=\int\d^3y \sqrt{h}\,\big[-\ca^a(\nabla_{[a}X_{b]}\Delta_Y\ca^{\dagger b}+X^b\nabla_b\Delta_Y\ca_a^\dagger)-\hc\big]\nn\\
&=-\int\d^3y \sqrt{h}\,\big[\ca^a\Delta_X\Delta_Y\ca_a^\dagger-\ca^{\dagger a}\Delta_X\Delta_Y\ca_a\big].
\end{align}
It follows that
\begin{align}
[\cm_X,\cm_Y]&=-m^2\int\d^3y \sqrt{h}\,\big[\ca^a[\Delta_Y,\Delta_X]\ca_a^\dagger-\ca^{\dagger a}[\Delta_Y,\Delta_X]\ca_a\big]\\
&=m^2\int\d^3y \sqrt{h}\,\big[\ca^a\Delta_{[X,Y]}\ca^\dagger_a-\ca^{\dagger a}\Delta_{[X,Y]}\ca^\dagger_a-o_{ab}(\ca^a\ca^{\dagger b}-\ca^{\dagger a}\ca^b)\big]\nn
\end{align}
which directly leads to \eq{MXMY1}. 

\paragraph{New operator.}
From \eq{MXMY1}, we define a new operator
\begin{align}
\cs_{s}&=\i m^2\int \d^3y \sqrt{h}\,s_{ab}(\ca^a\ca^{\dagger b}-\ca^{\dagger a}\ca^b)\nn\\
&=-2\i m^2\int \d^3y \sqrt{h}\,s_{ab}\ca^{\dagger a}\ca^b,\label{Sodef1}
\end{align}
where $s_{ab}(y)$ is an antisymmetric tensor. We first consider the action on the physical fields
\begin{align}
[\i \cs_s,\ca_a]=-s_{ab}\ca^b\equiv\delta_s\ca_a \qaq [\i \cs_s,\ca^\dagger_a]=-s_{ab}\ca^{\dagger b}\equiv\delta_s\ca_a^\dagger.\label{sgaa}
\end{align}
With these results, \eq{Sodef1} can be recast to the same form as \eqref{rewrite Tf Mx s1}, i.e.,
\begin{align}
\cs_s=-\i m^2\int \d^3y \sqrt{h}\,\big[\ca^a\delta_s\ca_a^\dagger-\ca^{\dagger a}\delta_s\ca_a\big],
\end{align}
We compute the commutator between two such operators
\begin{align}
[\cs_{s_1},\cs_{s_2}]=\i \cs_{s_{12}}\quad\qwq\quad (s_{12})_{ab}=(s_2)_{ac}(s_1)^c_{\ b}-(s_1\leftrightarrow s_2).
\end{align}

\paragraph{Extended charge algebra.}
With the operators $\mathcal T_f,\mathcal M_X$ and $\mathcal S_s$, the whole charge algebra can be worked out
\begin{subequations}\label{chargealgebrae}
\begin{align}
[\ct_{f_1},\ct_{f_2}]&=0,\\
[\ct_f,\cm_X]&=-\i\ct_{X(f)},\\
[\cm_X,\cm_Y]&=\i\cm_{[X,Y]}+\i\cs_{o(X,Y)},\label{intMM}\\
[\ct_f,\cs_s]&=0,\\
[\cm_X,\cs_s]&=\i\cs_{p(X,s)},\label{intMS}\\
[\cs_{s_1},\cs_{s_2}]&=-\i \cs_{[s_1,s_2]},\label{intSS}
\end{align}
\end{subequations}
where we have defined
\begin{align}
p_{ab}(X,s)=X^c\nabla_cs_{ab}-(s_{a}{}^c\nabla_{[c}X_{b]}-s_{b}{}^c\nabla_{[c}X_{a]})=X^c\nabla_cs_{ab}-\frac{1}{2}[s,\d X]_{ab},
\end{align}
and the bracket between two forms has the natural meaning, e.g.,
\begin{align}
[s_1,s_2]_{ab}=(s_1)_{ac}(s_2)^c{}_b-(s_1)_{bc}(s_2)^c{}_a.
\end{align}
It is interesting to find that the algebra has a similar structure to the intertwined Carrollian diffeomorphism\footnote{Carrollian diffeomorphism preserves the null structure of a Carrollian manifold \cite{Ciambelli:2018xat} and the intertwined Carrollian diffeomorphism indicates the inclusion of the superduality transformation. } in general dimensions \cite{Liu:2024rvz}. In section \ref{sd}, we will show that the new operator $\cS_s$ is a spin operator. Since the algebra generated by $\cM_X$ and $\cS_s$ is a deformation of the magnetic diffeomorphism by the spin operator, we will call the sub-algebra made up of \eqref{intMM}, \eqref{intMS}, and \eqref{intSS} as intertwined magnetic diffeomorphism on $H^3$ and denote it as $\text{IMDiff}(H^3)$. Correspondingly,  we will denote \eqref{chargealgebrae} as
\begin{align}
\text{IMDiff}(H^3)\ltimes C^\infty(H^3).
\end{align}

\subsection{Massive Fierz-Pauli field}
In this subsection, we consider the charge algebra obtained for massive spin-2 fields.

\paragraph{Charges.} The energy and angular momentum densities are 
\begin{subequations}
\begin{align}
T(y)&=2m^3 \cH^\dagger_{ab}\cH^{ab},\\
M_a(y)&=-\i m^2P^{abcdef}(\cH_{bc}\nabla_d \cH^\dagger_{ef}-\cH^\dagger_{bc}\nabla_d\cH_{ef}),
\end{align} 
\end{subequations}
where the rank $6$ tensor $P^{abcdef}$ can be written as 
\begin{align}
h^{ab}h^{ce}h^{df}+h^{ac}h^{be}h^{df}+h^{ad}h^{be}h^{cf}-h^{ae}h^{bd}h^{cf}-h^{af}h^{be}h^{cd}.
\end{align} 
It can be transformed to be symmetric under the exchange of indices $b\leftrightarrow c$ or $e\leftrightarrow f$ since $\cH_{bc}$ is symmetric. In other words, we can impose
\begin{align}
P^{abcdef}=P^{acbdef}=P^{abcdfe}=P^{acbdfe},
\end{align} 
and then get explicitly
\begin{align}
P^{abcdef}=&\,\frac{1}{2}(h^{ab}h^{ce}h^{df}+h^{ac}h^{be}h^{df}+h^{ab}h^{cf}d^{de}+h^{ac}h^{bf}h^{de})\nn\\&-\frac{1}{2}(h^{ae}h^{bd}h^{cf}+h^{af}h^{be}h^{cd}+h^{ae}h^{cd}h^{bf}+h^{af}h^{ce}h^{bd})\nn\\&+\frac{1}{2}(h^{ad}h^{be}h^{cf}+h^{ad}h^{bf}h^{ce}).
\end{align}
Then the corresponding charges on $i^+$ are 
\begin{align}
\mathcal T_f&=2m^3\int \d^3y \sqrt{h}\, f(y) :\cH^{\dagger ab}\cH_{ab}:,\\
\mathcal M_X&=-\i m^2\int \d^3y \sqrt{h}\, P^{abcdef}X_a:(\cH_{bc}\nabla_d \cH^\dagger_{ef}-\cH^\dagger_{bc}\nabla_d\cH_{ef}):.
\end{align} 
It is easy to find 
\begin{align}
[\i\mathcal T_f, \cH_{ab}]=\Delta_f \cH_{ab}\qaq [\i\mathcal M_X,\cH_{ab}]=\Delta_X \cH_{ab}, \label{varH}
\end{align} 
where 
\begin{subequations}
\begin{align}
\Delta_f \cH_{ab}&=-\i m f \cH_{ab},\\
\Delta_X\cH_{ab}&=X^c\nabla_c \cH_{ab}+\nabla_{[a}X_{c]}\cH^c_{\ b}+\nabla_{[b}X_{c]}\cH_a^{\ c}.
\end{align} 
\end{subequations}
To derive \eqref{varH}, we should use the identity 
\begin{align}
\frac{1}{2}P^{abcdef}\mathcal P_{efmn}\nabla_d X_a \cH_{bc}&=\nabla_{[m }X_{c]} \cH^c_{\ n}+\nabla_{[n }X_{c]} \cH^c_{\ m}
\end{align} 
for divergence-free $X^a$ and symmetric traceless $\cH_{ab}$. Another useful identity is 
\begin{align}
(P^{abcdef}+P^{aefdbc})=h^{ad}h^{be}h^{cf}+h^{ad}h^{bf}h^{ce}.
\end{align}

\paragraph{Charge algebra.} It is straightforward to compute
\begin{align}
\left([\Delta_X, \Delta_Y]-\Delta_{[X,Y]}\right)\cH_{ab}=-o_{ac}(X,Y)\cH^c_{\ b}-o_{bc}(X,Y)\cH^c_{\ a}.\label{MXMYOo}
\end{align} 
This indicates the appearance of a quadratic parity-odd operator 
\begin{align}
\cS_s=-\i m^2\int \d^3y \sqrt{h}\, Q_{abcd}\cH^{\dagger ab}\cH^{cd}
\end{align} 
where the tensor $Q_{abcd}$ is defined as
\begin{align}
Q_{abcd}=\frac{1}{2}(s_{ac}h_{bd}+s_{bc}h_{ad}-s_{ca}h_{bd}-s_{cb}h_{ad}).
\end{align} 
Similar to the spin 1 case, the tensor $s_{ab}$ is skew-symmetric.  Therefore, the operator $\mathcal S_s$ can be simplified to 
\begin{align}
\cS_s=-4\i m^2 \int \d^3 y\sqrt{h}\, s_{ab}\cH^{\dagger ac}\cH^b_{\ c}.\label{sodefs2}
\end{align} 
It is easy to prove 
\begin{align}
[\i\cS_s,\cH_{ab}]=-s_{ac}\cH^{c}_{\ b}-s_{bc}\cH^c_{\ a}
\end{align} 
which matches the right-hand side of \eqref{MXMYOo} for $s_{ab}=o_{ab}$. Then the charge algebra is precisely isomorphic to its partner \eqref{chargealgebrae} for the Proca theory.

\section{Spin density and charge}\label{sd}
In this section, we will discuss various properties of the emerging spin charge operators.
\subsection{Case of spin 1}
In \eqref{Sodef1}, we have defined an operator in the Proca theory at $i^+$ whose density is\footnote{The insertion of the factor 2 in the spin density comes from the convention that the smeared operator $\cS_s$ is written as
\begin{align}
\cS_s=\frac{1}{2}\int \d^3y \sqrt{h}\, s_{ab}S^{ab}.\nn
\end{align}}
\begin{align}
S_{ab}(y)=-2\i m^2(\ca^{\dagger}_a\ca_b-\ca^{\dagger}_b\ca_a),
\end{align} 
which is antisymmetric and Hermitian
\begin{align}
S_{ab}=-S_{ba},\quad S_{ab}^\dagger=S_{ab}.
\end{align} 
Utilizing the Levi-Civita tensor of $H^3$, we can define a pseudo-vector 
\begin{align}
S^a=\frac{1}{2}\epsilon^{abc}S_{bc}.
\end{align} 

Now we switch to the locally flat frame by virtue of vielbein $e^i_a$
\begin{align}
e^i_a e^j_bh^{ab}=\delta^{ab},\quad e^i_a e^j_b\delta_{ij}=h_{ab}.
\end{align} 
Namely, introduce three independent operators 
\begin{align}
S^i=e^i_a S^a=-2\i m^2\epsilon^{ijk}\cA^\dagger_j\cA_k,
\end{align} 
where $\ca_i=e_i^a\ca_a$ satisfies
\begin{align}
[\ca_i(y),\ca_j^\dagger(y')]
&=\frac{1}{2m^2}\delta_{ij}\delta^{(3)}(y-y').
\end{align}
We can define an integral
\begin{align}
\cs_i=\int \d^3y \sqrt{h}\,S_i(y)=-2\i m^2\int \d^3 y \sqrt{h}\, \epsilon_{ijk}\ca_j^\dagger \ca_k,\label{si}
\end{align} 
which corresponds to the smeared operator $\cs_s$ with the choice 
\begin{align}
s_{ab}=\epsilon_{ijk}e^j_a e^k_b.
\end{align}

It is easy to compute
\begin{align}
[\cs_{i},\ca_j]=\i\epsilon_{ijk}\ca_k \qaq [\cs_{i},\ca^\dagger_j]=\i\epsilon_{ijk}\ca^\dagger_k,
\end{align}
as well as
\begin{align}
[\cs_{i},\cs_{j}]=\i \cs_{ij}=\i\epsilon_{ijk}\cs_{k}.
\end{align}
We have recovered the commutation relation for (spin) angular momentum. Therefore, $\cs_i$ are three independent spin operators for the Proca field and $S^a$ may be interpreted as the spin density operator at $i^+$.

\paragraph{Mode expansion.}
To further verify that $\cs_i$ is the spin charge, we substitute \eqref{camexp} into \eq{Sodef1}
\begin{align}
\cs_i&=-\frac{\i m^2}{2(2\pi)^3}\sum_{\l,\l'}\int \d^3y \sqrt{h}\,\epsilon^{ijk}e_j^a e_k^b S^\m_aS^\n_b\epsilon^{*\l}_\m\epsilon^{\l'}_\n a^\dagger_{\l}a_{\l'}\nn\\
&=-\frac{\i m^2}{2(2\pi)^3}\sum_{\l,\l'}\int \d^3y \sqrt{h}\,\epsilon^{ijk}\epsilon_j^{*\lambda}\epsilon_k^{\lambda'}a_{\lambda}^\dagger a_{\lambda'},
\end{align} 
where we have defined $\epsilon_j^\lambda=e_j^a S_a^\mu \epsilon_\mu^{\lambda}$ and thus the orthogonality and completeness relations become
\begin{align}
\epsilon_j^{*\lambda}(\bm p)\epsilon^{j\lambda'}(\bm p)=\delta^{\lambda\lambda'},\quad \sum_{\lambda}\epsilon^{*\lambda}_j(\bm p)\epsilon_{k\lambda}(\bm p)=\delta_{jk}.\label{orthocomp}
\end{align} 
Using the relation
\begin{align}
\d^{3}p=|\det \p_ap^i|\d^3y=m^3\cosh\r\sqrt{h}\d^3y=m^2\omega_{\bm p}\sqrt{h}\d^3y
\end{align} 
and introducing the spin matrix
\begin{align}
S^{\l,\l'}_i{=\epsilon_{ijk}\epsilon^{*\l}_j\epsilon^{\l'}_k},
\end{align}
the operator $\cs_i$ can be converted to 
\begin{align}
\cs_i=-\i\sum_{\l,\l'}\int \frac{\d^3p}{(2\pi)^3}\frac{1}{2\omega_{\bm p}}\,S^{\l,\l'}_ia^\dagger_{\l}a_{\l'}.
\end{align}
One can check that in this form, $\cs_i$ indeed satisfies the $\so(3)$ algebra 
\begin{align}
[\cs_i,\cs_j]&=-\sum_{\l_1,\cdots,\l_4}\int \frac{\d^3p}{(2\pi)^3}\frac{1}{2\omega_{\bm p}}\int \frac{\d^3q}{(2\pi)^3}\frac{1}{2\omega_{\bm q}}S^{\l_1,\l_2}_iS^{\l_3,\l_4}_j[a^\dagger_{\l_1}(\bm p)a_{\l_2}(\bm p),a^\dagger_{\l_3}(\bm q)a_{\l_4}(\bm q)]\nn\\
&=-\sum_{\l_1,\cdots,\l_4}\int \frac{\d^3p}{(2\pi)^3}\frac{1}{2\omega_{\bm p}}S^{\l_1,\l_2}_iS^{\l_3,\l_4}_j(\delta_{\l_2\l_3}a^\dagger_{\l_1}a_{\l_4}-\delta_{\l_1\l_4}a^\dagger_{\l_3}a_{\l_2})\nn\\
&=-\sum_{\l,\l'}\int \frac{\d^3p}{(2\pi)^3}\frac{1}{2\omega_{\bm p}}\ep_{ikl}\ep_{jmn}(\delta_{lm}\ep^{*\l}_k\ep^{\l'}_n -\delta_{kn}\ep^{*\l}_m\ep^{\l'}_l )a^\dagger_{\l}a_{\l'}\nn\\
&=\sum_{\l,\l'}\int \frac{\d^3p}{(2\pi)^3}\frac{1}{2\omega_{\bm p}}\delta^{m}_i\delta^{n}_j(\epsilon^{*\l}_m\epsilon^{\l'}_n-\epsilon^{*\l}_n\epsilon^{\l'}_m)a^\dagger_{\l}a_{\l'}\nn\\
&=\i\ep_{ijk}\times(-\i)\sum_{\l,\l'}\int \frac{\d^3p}{(2\pi)^3}\frac{1}{2\omega_{\bm p}}\ep_{klm}\epsilon^{*\l}_l\epsilon^{\l'}_ma^\dagger_{\l}a_{\l'}\nn\\
&=\i\ep_{ijk} \cs_k.
\end{align}

\paragraph{As a Noether charge.}
As a matter of fact, we can use Noether's theorem to derive the angular momentum current and decompose it into the summation of orbital and spin angular momentum current. Under an infinitesimal Lorentz transformation $\L_\m{}^\n=\delta_\m^\n+\delta\omega_\m{}^\n$, we find
\begin{align}
\delta_\omega A_\m(x)&=A'_\m(x)-A_\m(x)\nn\\
&=-\delta\omega_\m{}^\n A_\n(x)-\delta\omega^\r{}_\n x^\n\p_\r A_\m(x)+O(\delta\omega^2).
\end{align} 
The second part is related to the orbital angular momentum and the variation for intrinsic spin is
\begin{align}
\delta_\omega^{\rm spin}A_\m=-\delta\omega_\m{}^\n A_\n.
\end{align}
Obviously, this corresponds to the boundary variation \eq{sgaa}. Using Noether's procedure, one can derive the angular momentum current
\begin{align}
J^\mu_{\delta\omega}=\cdots+S^\mu_{\delta\omega}
\end{align} 
where we omit the orbital part. The spin part reads
\begin{align}
S_{\delta\omega}^\m=F^\mn\delta\omega_{\n\r}A^\r \qrq S^{\mn\r}=2F^{\m[\n}A^{\r]}.
\end{align}
Taking integration on $i^+$, one obtain
\begin{align}
\cS^\a{}_\b&=\int_{i^+}(\d^3x)_\t (F^{\t\a}A_{\b}-F^{\t}{}_\b A^{\a}).
\end{align}
Note that the parameter $\delta\omega_\mn$ is a dimensionless constant in the Cartesian frame, which implies $\delta\omega^\a{}_\b$ is of order $\t^0$ and $S^{\t\a}{}_\b$ will give rise to the appropriate (components of) spin charge. We are interested in the pure spatial component
\begin{align}
\cs^{a}{}_b=-2\i m^2\int\d^3y \sqrt{h}\,(\ca^{\dagger a}\ca_b-\ca^a\ca^\dagger_b)
\end{align}
which is the spin angular momentum and of course, a special case of our extended spin charge \eqref{Sodef1}. 

\paragraph{Decomposition of $\cm_X$.}
A symmetric stress tensor can also be obtained from canonical Noether's formalism along with the Belinfante method, which implies that the angular momentum derived from this stress tensor already included the contribution from spin. Therefore, we may decompose the charge operator $\cM_{X}$ \eqref{MXs=1} through integration by parts (and discarding the boundary term)
\begin{align}
\cm_X=-\i m^2\int\d^3y \sqrt{h}\, \big[\nabla_aX_b(\ca^a\ca^{\dagger b}-\ca^{\dagger a}\ca^b)+X^a(\cA^b\nabla_a\ca^\dagger_b-\cA^{\dagger b}\nabla_a\ca_b)\big].\label{MXdecomp}
\end{align}
It is obvious that the first term is the contribution from spin, while the second term comes from the ``orbital'' part. Namely, we may decompose it as 
\begin{align}
\cM_X=\co_X -\frac{1}{2}\cS_{\d X}
\end{align} 
where the ``orbital'' part is 
\begin{align}
\co_X=-\i m^2\int \d^3y \sqrt{h}\,X^a(\cA^b\nabla_a\cA^\dagger_b-\cA^{\dagger b}\nabla_a\cA_b).
\end{align}
The commutators between the ``orbital'' and spin operators are
\begin{subequations}\label{orbit}
\begin{align}
[\co_X,\co_Y]&=\i\co_{[X,Y]}+\i\cS_{q(X,Y)},\\
[\co_X,\cS_s]&=\i\cS_{X^c\nabla_c s_{ab}},\\
[\cS_{s_1},\cS_{s_2}]&=-\i\cS_{[s_1,s_2]}.
\end{align}
\end{subequations}
The antisymmetric tensor $q(X,Y)$ is defined as
\begin{align}
q(X,Y)&=o(X,Y)-\frac{1}{2}\d[X,Y]+\frac{1}{2}p(X,\d Y)-\frac{1}{2}p(Y,\d X)-\frac{1}{4}[\d X,\d Y].
\end{align} 
It is interesting to find 
\begin{align}
q_{ab}(X,Y)=-R_{abcd}X^c Y^d
\end{align} 
where $R_{abcd}$ is the Riemann tensor of $H^3$.  Unfortunately, the ``orbital'' part does not form a representation of the rotation group even for Killing vectors $X$ and $Y$ due to the anomalous Riemann tensor term. A similar algebra has been found in \cite{Liu:2024rvz}.

\subsection{Case of spin 2}
For the massive spin 2 field, the spin density operator would be 
\begin{align}
S^a=\frac{1}{2}\epsilon^{abc}S_{bc} \qwq S_{ab}=-4\i m^2(\cH_{ac}^\dagger \cH_b^{c}-\cH_{bc}^\dagger \cH_a^{c}).
\end{align} 
It can be switched to the local flat frame 
\begin{align}
S^i=e_a^i S^a=-4\i m^2\epsilon^{ijk}\cH_{jl}^\dagger \cH_{k}^{l} \qwq \cH_{ij}=e^a_i e^b_j \cH_{ab},
\end{align}
and the corresponding spin charge is
\begin{align}
\cS_i=\int \d^3 y\sqrt{h}\, S_i=-4\i m^2\epsilon_{ijk}\int \d^3 y \sqrt{h}\, \cH^{\dagger jl}\cH^k_{l}.
\end{align} 
We find the commutator 
\begin{align}
[\cS_i,\cH_{jk}]=\i(\epsilon_{ijl}\cH_{kl}+\epsilon_{ikl}\cH_{jl})
\end{align} 
and reproduce the $\so(3)$ algebra 
\begin{align}
[\cS_i,\cS_j]=\i\epsilon_{ijk}\cS_k.
\end{align}

One can also analyze the mode expansion of the spin charge for the massive Fierz-Pauli field using the same method as the Proca field, although the expression will be more complicated. We will not do so here. Instead, we can compute the spin charge from the Noether current \eq{smnr}
\begin{align}
\cs^a{}_b&=\int(\d^3x)_\t\, t_\m S_\n^a S_{\r b}\Big[4\p^\m H^{\s[\n}H^{\r]}_\s-2(\p^{\n}H^{\m\s}H^{\r}_\s+\p^\s H^{\mn}H^{\r}_\s-\n\leftrightarrow\r)\Big]\nn\\
&=4\i m^2\int \d^3y\sqrt{h}\,(\ch^{\dagger ac}\ch_{bc}-\ch^{ac}\ch^\dagger_{bc}),
\end{align}
which is a special case of our extended spin charge \eqref{sodefs2}. Moreover, let us consider the decomposition of the charge $\cm_X$. After integrating by parts, we obtain
\begin{align}
\cm_X&=-\i m^2\int\d^3y \sqrt{h}\,\big[2\nabla_aX_b\ch^{ac}\ch^{\dagger b}_c +X^a\ch^{bc}\nabla_a\ch^\dagger_{bc}-\hc\big]\\
&\equiv\co_X-\frac{1}{2}\cs_{\d X},
\end{align}
where the ``orbital'' part is
\begin{align}
\co_X&=-\i m^2\int\d^3y \sqrt{h}\,X^a(\ch^{bc}\nabla_a\ch^\dagger_{bc}-\ch^{\dagger bc}\nabla_a\ch_{bc}).
\end{align}
This decomposition is totally the same as the Proca case, and we know the extended charge algebras also coincide, so the commutators between these two operators are the same as \eqref{orbit}.

\section{Comparisons}\label{comp}
In this section, we will compare our extended charge algebra with various algebras found in the literature. 
\subsection{BMS algebra at $i^+$}
In the BMS group, there is a class of infinite-dimensional diffeomorphism called supertranslation. The standard derivation is to find the diffeomorphism preserving the Bondi gauge and asymptotic expansion of the dynamic metric. 

However, as a post hoc derivation, the leading order of the supertranslation vector field $\wt\xi_T$ can be obtained from extending the translation generator $\xi_c$. For example, in retarded coordinates $(u,r,x^A)$ we have
\begin{align}
\xi_c=c^\m\p_\m=c^\m(-n_\m \p_u+m_\m\p_r-r^{-1}Y^A_\m\p_A),
\end{align}
where $m_\m=(0,n_i)$. Noticing the following relations
\begin{align}
m_\m=-\frac{1}{2}\nabla_A\nabla^An_\m \qaq Y_\m^A=-\nabla^An_\m,
\end{align}
it is natural to extend the common factor $-c^\m n_\m$ to a general $T(\Omega)$ and we get
\begin{align}
\wt\xi_T=T\partial_u+\frac{1}{2}\nabla_A\nabla^A T\partial_r-\frac{1}{r}\nabla^A T\partial_A+\cdots,\label{xifscri}
\end{align}
which is exactly the leading order of the supertranslation vector field whose subleading orders rely on the dynamic components of the asymptotically flat metric. Moreover, one can check \eqref{xifscri} is divergence-free
\begin{align}
    \nabla_\mu \wt\xi^\mu_T=0.\label{divf}
\end{align} 
The same logic applies to the Lorentz generator, from which one can get
\begin{align}
\wt\xi_Y=\frac{1}{2}u\nabla\cdot Y\partial_u-\frac{1}{2}r\nabla\cdot Y\partial_r+\frac{u }{4}\nabla^2\nabla\cdot Y\p_r+(Y^A-\frac{u}{2r}\nabla^A\nabla\cdot Y)\partial_A+\cdots\label{xiYscri}
\end{align}
which is the superrotation vector field in the sense of the generalized BMS group. This $\wt\xi_Y$ is also divergence-free
\begin{align}
\nabla_\mu\wt\xi^\mu_Y=0.\label{divY}
\end{align}
The divergence-free conditions \eqref{divf} and \eqref{divY} is not surprising since the $\wt\xi_{T,Y}$ are asymptotic Killing vectors that preserve the boundary fall-off conditions.

Now we follow the same method to extend the Poincar\'e generator near $i^+$. From \eqref{tipm}, we find
\begin{align}
c^\m\p_\m=c^\m(t_\m\p_\t-\frac{1}{\t}D^at_\m\p_a)\label{transf}
\end{align}
which can be naturally generalized to
\begin{align}
\xi_f=f(y)\p_\t-\frac{1}{\t}D^af(y)\p_a.
\end{align}
Demanding $\xi_f$ to be divergence-free, we obtain
\begin{align}
\nabla_\mu\xi^\m_f=0 \qrq (\nabla_a\nabla^a-3)f=0\label{transf2}
\end{align}
which agrees with the BMS-like supertranslation found in \cite{Chakraborty:2021sbc,Compere:2023qoa}. An asymptotically flat spacetime near $i^+$ admits the fall-off conditions \cite{Compere:2023qoa}\footnote{The asymptotic expansion can be obtained from the one near $i^0$ by an analytic continuation \cite{beig_einsteins_1982,beig1984integration}.}
\begin{align}
\d s^2&=[-1-\frac{2\sigma}{\tau}-\frac{\sigma^2}{\tau^2}+o(\tau^{-2})]\d\tau^2+o(\tau^{-2})\tau \d\tau \d y^a\nn\\
&\quad +\tau^2[h_{ab}+\frac{k_{ab}-2\sigma h_{ab}}{\tau}+\frac{\log\tau}{\tau^2}i_{ab}+\frac{1}{\t^2}j_{ab}+o(\tau^{-2})]\d y^a \d y^b
\end{align} 
where $\sigma,k_{ab}$ are first order and $i_{ab},j_{ab}$ are second order fields at $i^+$. The field $\sigma$ is determined by the source in the bulk and vanishes at large $\rho$, which could be used to remove a logarithmic translation degree of freedom. On the other hand, the field $k_{ab}$ is assumed to be a pure gauge. The asymptotic symmetric group is generated by 
\begin{align}
\xi_{f,X}=\left(f-\frac{1}{\tau}(\sigma f+\nabla_a\sigma\nabla^af)+o(\tau^{-1})\right)\partial_\tau+\left(X^a-\frac{1}{\tau}\nabla^a f+o(\tau^{-1})\right)\partial_a
\end{align} 
where $f$ obeys the equation 
\begin{align}
(\nabla_a\nabla^a-3) f=0\label{nablaf}
\end{align} 
following from the traceless condition of $k_{ab}$. The vector $X^a$ is a KV of $H^3$ so that it preserves the boundary metric $h_{ab}$
\begin{align}
\nabla_{(a} X_{b)}=0.\label{KVX}
\end{align}
In a flat spacetime, $\sigma=0$ and the asymptotic Killing vector has the same form as \eqref{liftfX}. Note that the supertranslation equation \eqref{nablaf} could be obtained by extending the translation generator and imposing the divergence-free condition, as shown from \eqref{transf} to \eqref{transf2}. Therefore, we conclude that our charge algebra is reduced to BMS algebra at $i^+$ under the condition \eqref{nablaf} and \eqref{KVX}.

However, these conditions are not necessary in our framework since the general vector $\xi_{f,X}$ leads to a charge whose action on the fundamental field agrees with the covariant variation. Notice that the extended charge algebra for these more general vectors does not necessarily generate asymptotic symmetry.

\subsection{Generalized BMS algebra at $i^+$}\label{gmbsi+}
In \cite{Campiglia:2015kxa}, the authors derived the generalized BMS vector fields at $i^+$
as residual large gauge transformations that preserve the de Donder gauge and certain fall-off conditions.  The vector field is found to be 
\begin{subequations}
\begin{align}
\xi_{f}&=[f+o(1)]\partial_\tau-[\frac{1}{\tau}\nabla^af+o(\tau^{-1})]\partial_a,\\
\xi_X&=o(1)\partial_\tau+[X^a+o(1)]\partial_a
\end{align} 
\end{subequations}
where the supertranslation function $f$ still obeys the equation \eqref{nablaf} and the superrotation vector field $X^a$ satisfies
\begin{align}
(\nabla_a\nabla^a-2)X^b=0 \qaq \nabla_aX^a=0\label{generalX}
\end{align} 
instead of the Killing equations. 

It seems that one can impose the same conditions for $f$ and $X^a$ in our case. However, the situation is much more involved. Although we may write the commutator \eqref{TfMXscalar} as 
\begin{align}
[\text{supertranslation},\text{superrotation}]=\text{supertranslation},
\end{align} 
it is necessary to check whether $X^a\nabla_af$ satisfies the condition of supertranslation \eqref{nablaf}. Unfortunately, we find 
\begin{align}
(\nabla^2-3)(X^a\nabla_a f)=(\nabla^2-2)X^a\nabla_a f+2\nabla^aX^b\nabla_a\nabla_b f+X^a\nabla_a(\nabla^2-3)f,
\end{align} 
where we have used the identities
\begin{align}
[\nabla_a,\nabla_b]V^c=R^c{}_{dab}V^d \qaq R_{ab}=-2h_{ab}.
\end{align}
The first and third term on the right-hand side vanishes via the conditions \eqref{nablaf} and \eqref{generalX}. However, the second term survives except that $X^a$ is a Killing vector. One can also check that $[X,Y]$ does not necessarily satisfy the first constraint equation of \eqref{generalX}. We conclude that the generalized BMS algebra is not a sub-algebra of our result. There are two ways to find a consistent algebra:
\begin{enumerate}
    \item We can restrict $X^a$ to be a Killing vector and then the algebra becomes the standard BMS algebra.
    \item We can relax $X^a$ such that only the divergence free condition $\nabla_a X^a=0$ is satisfied. The resulting algebra is $\text{MDiff}(H^3)\ltimes C^\infty(H^3)$.
\end{enumerate}
Note that the closure of the generalized BMS algebra at timelike infinity has been discussed in \cite{AH:2020rfq}. In their formulation, the commutator between a supertranslation (superrotation) vector field and a superrotation vector field is still a supertranslation (superrotation) vector field since the Lie bracket of two vectors has been replaced by the modified Lie bracket \cite{Barnich:2011mi}. In our case, we find that the commutators between covariant variations agree with the charge algebra. Therefore, we do not try to use their modified Lie bracket in our work. 

\subsection{Flux algebra at $\mathcal I^+$}
It is interesting to find the charge algebra \eqref{chargealgebrae} has exactly the same form as (2.29) in \cite{Liu:2024rvz} after taking $\dot f=0$  in the latter case. The second algebra is the flux algebra for the intertwined generalized BMS group at future null infinity, which was reproduced here by the charges. Some correspondences between the charge algebra at $i^+$ and the flux algebra at $\mathcal I^+$  are collected in table \ref{table_ti_scri}. Note that in four dimensions, the commutator of the helicity flux operators in the latter algebra vanishes. However, the spin operator in the former algebra is non-Abelian and thus the commutator of two spin operators does not vanish. The non-Abelian structure follows from the massive representation of the Poincar\'e group. Note that one can lift the flux algebra at $\mathcal I^+$ to the five-dimensional spacetime, and then the helicity flux operators form a non-Abelian representation. Through replacing the parameters on $H^3$ by those on $S^3$, the charge algebra at $i^+$ in 4 dimensions is mapped to the (magnetic) flux algebra at $\mathcal I^+$ in 5 dimensions. At last, the helicity flux density 2-form in 4 dimensions is equivalent to a function $O(u,\Omega)$ since $O_{AB}$ is proportional to the Levi-Civita tensor $\epsilon_{AB}$ on $S^2$. For the spin density at $i^+$, the same thing happens in 3 dimensions which is shown in appendix \ref{secddim}.

\begin{table}[htbp]
\centering\renewcommand{\arraystretch}{1.25}
\begin{tabular}{|c|c|c|}
\hline
& Timelike infinity & Null infinity  \\ \hline
Manifold & $i^+\simeq H^{3}$ & $\ci^+\simeq\R\times S^{2}$\\ \hline
Algebra (scalar field)&  $\text{MDiff}(H^3)\ltimes C^{\infty}(H^3)$& $\text{Diff}(S^2)\ltimes C^\infty(S^2)$\\\hline
Algebra (spinning fields)&$\text{IMDiff}(H^3)\ltimes C^{\infty}(H^3)$&$\text{IDiff}(S^2)\ltimes C^\infty(S^2)$\\\hline
Emerging operator& Spin density $S_{ab}$  & Helicity flux density $O_{AB}$ \\ \hline
Supertranslation & $f(y)$ with $(\nabla_a\nabla^a-3)f=0$ & Smooth $T(\Omega)$ on $S^2$\\ \hline
\multirow{2}{*}{Lorentz transformation} & $X^a$, KV on $H^{3}$ & $Y^A$, CKV on $S^{2}$\\ \cline{2-3}
& $\nabla_{(a}X_{b)}=0$ &  $2\nabla_{(A}Y_{B)}=\g_{AB}\nabla\cdot Y$\\ \hline
Superrotation & $X^a$ with $\nabla_aX^a=0$ & Smooth $Y^A(\Omega)$\\ \hline
\end{tabular}
\caption{We list some correspondences between $i^+$ and $\ci^+$ of asymptotically Minkowski spacetime in 4 dimensions. We here call what the divergence-free $X^a$ generates ``superrotation'' to complete the list. This is only justified by the fact that its covariant variation agrees with the quantum commutator. The superrotation in the sense of asymptotic symmetry analysis still needs more exploration.\protect\footnotemark}
\label{table_ti_scri}
\end{table}
\footnotetext{We thank Geoffrey Comp\`ere for useful comments on the superrotation at timelike and spatial infinities.}

\subsection{More comparisons}
It is stated in \cite{Laddha:2022nmj} that the reduction of massive fields to a hyperboloid conformal to $i^+$ is satisfactory for the purpose of defining the S-matrix, but not suitable from the view of holography. Therefore, the authors develop a novel asymptotic description which basically extrapolates the massive fields to (the blow-up of) spatial infinity $i^0$ since it is a timelike hypersurface and thus the boundary theory can have interaction. This is indeed more like the usual AdS/CFT pattern where the boundary is timelike and some CFT lives on it. 

{However, we think that holography should have a more extensive meaning.} If we want to construct a holography in the asymptotically flat spacetime, then we can not require it to be the same as in the AdS space since many things are different. For instance, what we have is an infinite boundary that is made up of five parts. Timelike and null infinity are related to massive and massless particles, respectively, while spatial infinity is of less direct interest since it is causally separated from the finite region where we live and the interaction occurs. Due to the existence of the null boundary and the leaky boundary condition for gravitational radiation (see \cite{Donnay:2023mrd} and references therein), we can not ``put gravity in a box'' in the asymptotically flat space like in the AdS space and therefore, we have to address the holography principle beyond its usual set-up. What we aim to do is to encode the physics of (asymptotically) flat spacetime into a theory living at the boundary. In this setting, many successes are achieved, e.g., the establishment of the infrared triangle \cite{Strominger:2013jfa,He:2014laa,Strominger:2014pwa,Strominger:2017zoo} and the proposal of celestial/Carrollian holography which tries to represent the bulk scattering amplitudes by the correlators on the celestial sphere/null infinity \cite{Pasterski:2016qvg,Pasterski:2021raf,Donnay:2022aba,Bagchi:2022emh,Donnay:2022wvx,Donnay:2023mrd}. Following the same spirit, we explore the boundary massive fields\footnote{To highlight the property of massive fields and for convenience, we do not write out the gravity part which is of course explored separately in the literature.} which naturally live on the timelike infinity and can be seen as the initial and final states for the massive scattering. Along this road, the next step is to investigate the boundary amplitudes for massive scattering and scattering with both massive and massless particles, which will be explicitly illustrated in section \ref{conc}.

On the other hand, the spatial infinity has a  dual description with the timelike infinity, and they can be related through a simple coordinate transformation
\begin{align}
\hat\r=\i\t \qaq \hat\t=\r-\frac{\i\pi}{2}
\end{align}
where $\hat\r=\sqrt{r^2-t^2}$ and $\hat\t={\rm arctanh}\,(t/r)$ are coordinates suitable for describing the spatial infinity. The unit 3-dimensional hyperboloid for timelike infinity may be relabeled by $\ch^+$ which is known as Euclidean AdS$_3$, while $i^0$ is also conformal to a hyperboloid denoted by $\ch^0$, i.e., a Lorentzian dS$_3$. These are given for example, in section 2 of \cite{Compere:2023qoa} which contains more discussion on relating two regions in the view of gravity. As a result, the massive field in \cite{Laddha:2022nmj} shares a similar fall-off\footnote{Near $i^0$, the branch with fall-off $\hat\rho^{-3/2}\e^{m\hat\rho}$ is ruled out since the field blows up.} as ours ($\hat\rho^{-3/2}\e^{-m\hat\rho}$ vs. $\t^{-3/2}\e^{\pm\i m\t}$), which is known by the authors since they also reviewed the method we use. The previous arguments indicate that an Euclidean theory on $\mathcal H^+$ may be switched to a Lorentzian theory on $\mathcal H^0$ through analytic continuation. At last, it is always good to develop new methods as either alternatives or supplements.

\paragraph{Similar algebra.} The author of \cite{Schwarz:2022dqf} has found a similar deformation of the diffeomorphism algebra that also involves a spin operator, but he argued that it should be forbidden since the conservation of conformal spin leads to conservation of helicity which is definitely wrong in the physical process. We have noticed this paper in \cite{Liu:2023qtr} and commented in the conclusion part. Now we give a more detailed comparison:
\begin{itemize}
\item In our methods, the spin operator (or helicity flux at $\ci^+$) naturally emerges from the superrotation commutators, both classically (commutator of covariant variation) and quantum mechanically (quantum commutator of operators). Our operator is a smeared integration of the local density over hypersurfaces.
\item In \cite{Schwarz:2022dqf}, the introduction of the spin operator is to solve the problem of violating the Jacobi identity of $J\bar J\Phi$, where $J$ and $\bar J$ are generators for diffeomorphism and $\Phi$ is a conformal operator. After adding the spin operator $S$ in the commutator $[J,\bar J]$, the structure of their algebra is equivalent to ours \eqref{chargealgebrae} with $\ct_f$ excluded.
\end{itemize}
At $\ci^+$, the flux is not a conserved quantity since we have a leaky boundary condition, and there is radiation across the boundary. At $i^+$, the conserved quantities are the Poincar\'e charges, while the extended charges are not required to be conserved. In both cases, the helicity fluxes/spin charges are smeared composite operators integrated over the boundaries, and there is no reason to demand a conservation law for them. In this sense, our algebra is not an exact symmetry algebra for the matter field unless we restrict it to the Poincar\'e sub-algebra. Correspondingly, we cannot rule out the diffeomorphism algebra and helicity fluxes/spin charges by the argument of helicity {non-}conservation in a physical process.   

\section{Conclusion and discussion}\label{conc}
In this work, we have expanded the massive fields near $i^+$ and treated the coefficients in the expansion series as boundary fields at $H^3$. The fundamental fields are free and encode the outgoing data for a scattering process, which can be used to realize the Poincar\'e algebra at $H^3$. By extending the Poincar\'e charges, we could find a larger algebra which is denoted as $\text{MDiff}(H^3)\ltimes C^\infty(H^3)$. Here, $\text{MDiff}(H^3)$ means the magnetic diffeomorphisms that are generated by divergence-free vectors on $H^3$. The Abelian ideal of the algebra is composed of the smooth functions of $H^3$. For the spinning fields, one should include an additional spin charge operator to close the algebra. We have discussed how to reduce the algebra to the BMS algebra and also compared it with the Carrollian diffeomorphism. There are various problems that deserve study in the future.

\begin{figure}[tbp]
\centering
\subfloat[Massive $m\to n$ scattering]{\scalebox{0.8}{
\begin{tikzpicture}
  \draw[thick,blue,dashed] plot[smooth,tension=0.7] coordinates{(2,3) (0,2.4) (-2,3)};
  \draw[thick,blue,dashed] plot[smooth,tension=0.7] coordinates{(2,-3) (0,-2.4) (-2,-3)};
  \draw[thick,red,dashed] plot[smooth,tension=0.7] coordinates{(3,1.25) (2.5,0) (3,-1.25)};
  \draw[thick,red,dashed] plot[smooth,tension=0.7] coordinates{(-3,1.25) (-2.5,0) (-3,-1.25)};
  \draw[thick,brown,dashed] (3,1.25)--(2,3);
  \draw[thick,brown,dashed] (3,-1.25)--(2,-3);
  \draw[thick,brown,dashed] (-3,1.25)--(-2,3);
  \draw[thick,brown,dashed] (-3,-1.25)--(-2,-3);
  \fill (2,3) circle (1.5pt);
  \fill (-2,3) circle (1.5pt);
  \fill (2,-3) circle (1.5pt);
  \fill (-2,-3) circle (1.5pt);
  \fill (3,1.25) circle (1.5pt);
  \fill (3,-1.25) circle (1.5pt);
  \fill (-3,1.25) circle (1.5pt);
  \fill (-3,-1.25) circle (1.5pt);
  \node at (0,2.8) {$i^+$};
  \node at (3,2.25) {$\ci^+$};
  \node at (3,-2.1) {$\ci^-$};
  \node at (2.8,0) {$i^0$};
  \node at (0,-2.8) {$i^-$};
  \node at (-3,2.25) {$\ci^+$};
  \node at (-3,-2.1) {$\ci^-$};
  \node at (-2.8,0) {$i^0$};
  \node at (-1.6,-2.6) {$y_1$};
  \node at (0,-1.5) {$\cdots$};
  \node at (1.7,-2.6) {$y_m$};
  \node at (-1.75,2.4) {$y_{m+1}$};
  \node at (0,1.5) {$\cdots$};
  \node at (1.85,2.4) {$y_{m+n}$};
  \draw[mid arrow,thick, blue!60] (-1.4,-2.75)--(0,0);
  \draw[mid arrow,thick, blue!60] (1.4,-2.75)--(0,0);
  \draw[mid arrow,thick, blue!60] (0.2,0.38)--(1.5,2.78);
  \draw[mid arrow,thick, blue!60] (-0.2,0.38)--(-1.5,2.78);
  \filldraw[fill=gray,draw,thick] (0,0) circle (0.6);
\end{tikzpicture}
\label{mton}
}}
\hfil
\subfloat[$m_1=n_1=m_2=n_2=1$ scattering]{\scalebox{0.8}{
\begin{tikzpicture}
  \draw[thick,blue,dashed] plot[smooth,tension=0.7] coordinates{(2,3) (0,2.4) (-2,3)};
  \draw[thick,blue,dashed] plot[smooth,tension=0.7] coordinates{(2,-3) (0,-2.4) (-2,-3)};
  \draw[thick,red,dashed] plot[smooth,tension=0.7] coordinates{(3,1.25) (2.5,0) (3,-1.25)};
  \draw[thick,red,dashed] plot[smooth,tension=0.7] coordinates{(-3,1.25) (-2.5,0) (-3,-1.25)};
  \draw[thick,brown,dashed] (3,1.25)--(2,3);
  \draw[thick,brown,dashed] (3,-1.25)--(2,-3);
  \draw[thick,brown,dashed] (-3,1.25)--(-2,3);
  \draw[thick,brown,dashed] (-3,-1.25)--(-2,-3);
  \fill (2,3) circle (1.5pt);
  \fill (-2,3) circle (1.5pt);
  \fill (2,-3) circle (1.5pt);
  \fill (-2,-3) circle (1.5pt);
  \fill (3,1.25) circle (1.5pt);
  \fill (3,-1.25) circle (1.5pt);
  \fill (-3,1.25) circle (1.5pt);
  \fill (-3,-1.25) circle (1.5pt);
  \node at (0,2.8) {$i^+$};
  \node at (3,2.25) {$\ci^+$};
  \node at (3,-2.1) {$\ci^-$};
  \node at (2.8,0) {$i^0$};
  \node at (0,-2.8) {$i^-$};
  \node at (-3,2.25) {$\ci^+$};
  \node at (-3,-2.1) {$\ci^-$};
  \node at (-2.8,0) {$i^0$};
  \node at (2,1.85) {\scriptsize $(u_4,\Omega_4)$};
  \node at (-2,-1.9) {\scriptsize $(v_1,\Omega_1)$};
  \node at (-1.2,-2.4) {\scriptsize $y_2$};
  \node at (1.7,2.6) {\scriptsize $y_3$};
  \draw[mid arrow,thick, blue!60] (-1,-2.6)--(0,0);
  \draw[mid arrow,thick, blue!60] (0.2,0.38)--(1.5,2.78);
  \draw[snake it, thick, brown!60] (-2.7,-1.775)--(2.7,1.775);
  \filldraw[fill=gray,draw,thick] (0,0) circle (0.6);
\end{tikzpicture}\label{1111}
}}
\caption{In this figure, we show two kinds of scattering processes involving massive particles. In the left diagram, $m$ ingoing particles located originally at $y_1,\cdots,y_m$ become $n$ outgoing particles after scattering, and eventually arrive the location $y_{m+1},\cdots,y_{m+n}$ at $i^+$. In the right diagram, we depict a scattering process with input of $m_1=1$ massless particle coming from $\ci^-$ and $m_2=1$ massive particle coming from $i^-$, and the outputs are $n_1=1$ massless particle going to $\ci^+$ and $n_2=1$ massive particle going to $i^+$.}
\end{figure}
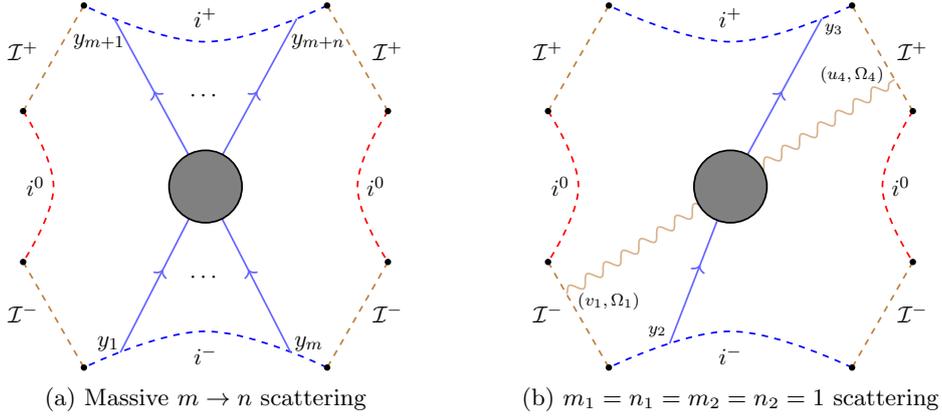

\begin{itemize}
\item \textbf{Null, spatial, and timelike infinities.} In an asymptotically flat spacetime, different asymptotic regions are connected through the joint corners. As for our concerns, the physics near the common boundary $i^+_\p=\ci^+_+=S^2$ is interesting. Although we can map the vector field near $\ci^+$ to $i^+$, the orders of large $r$ and large $\t$ will get mixed up. Only considering all the orders can give a match beyond the generator of Poincar\'e transformation.\footnote{As said in section \ref{gmbsi+}, the authors in \cite{Campiglia:2015kxa,AH:2020rfq} use the Green function to map the leading order of generalized BMS vector at $\ci^+$ to $i^+$ and consider the corresponding asymptotic analysis which leads to the mapped generalized BMS vector at $i^+$. Their analysis is different from the one of \cite{Compere:2023qoa}. Relaxing the boundary metric at $i^+$ may lead to a different ``superrotation'' than the mapped diffeomorphism coming from the celestial sphere. This is a point that needs further investigation.} It is interesting to explore whether we can find a natural way to compare the extended algebras for different fields and asymptotic regions. 

\item \textbf{Covariant variation.} In this paper, we find that the boundary covariant variation plays a key role in the agreement between the quantum commutator and classical variation for the spinning fields. The same phenomenon has been found at null infinity \cite{Liu:2023gwa,Liu:2023qtr,Liu:2023jnc}. The philosophy is that the extended transformation will change the bulk metric, and this change has a non-vanishing effect on the boundary physics. The calculation of the quantum commutator requires a fixed boundary metric. Therefore, we need to subtract the effect coming from the fluctuation of the boundary metric. It is natural to explore whether this logic applies to other hypersurfaces in general spacetime. Moreover, the introduction of covariant variation is not necessary to be limited to the boundary. As a matter of fact, we find that for example, the bulk covariant variation
\begin{align}
\Delta_{\xi_X}A_a=\cl_{\xi_X}A_a-\frac{1}{2}\cl_{\xi_X}\eta_{a\a}A^\a
\end{align}
gives the boundary covariant variation $\Delta_X\ca_a$ as its leading order. The same holds for all the cases with covariant variation we have found so far. It is interesting to investigate the geometric meaning\footnote{The bulk covariant variation can be seen as modifying both the Lie derivative and covariant derivative, whose definitions have natural geometric motivation.} and general property of the bulk covariant variation, such as the non-closure and Jacobi identity. A related paper is in progress.

\item \textbf{Partial Carrollian amplitude.}
The method used in this work is the same as \cite{Liu:2022mne} where the massless fields are extrapolated to future null infinity. In the latter case, the boundary field theory is supposed to be defined on the Carrollian manifold. In our case, the massive fields are reduced to $H^3$, the conformal boundary of $i^+$. Note that the boundary operator is exactly the annihilation or creation operator, i.e., \eqref{vpexp}, and thus we can use the boundary operators to define correlators on $H^3$
\begin{align}
\langle 0|\prod_{j=m+1}^{m+n} \varphi(y_j) \prod_{i=1}^m \varphi^{(-)\dagger}(y_i)|0\rangle\label{corre}
\end{align} 
where the superscript $(-)$ denotes the field at past timelike infinity. The argument $y_{i/j}$ is the inserted location of the corresponding fields. Note that the coordinate $y$ is also equivalent to the momentum of the outgoing/ingoing mode. We conclude that the correlator \eqref{corre} is exactly equivalent to the scattering amplitude in an $m\to n$ process (see figure \ref{mton}). Note that \eqref{corre} is the analog of the Carrollian correlator in the framework of bulk reduction \cite{Liu:2024nfc,Liu:2024llk,Li:2024kbo,Long:2025bfi}. Unlike the Carrollian amplitude \cite{Donnay:2022aba,Bagchi:2022emh,Mason:2023mti,Liu:2024nfc,Alday:2024yyj,Liu:2024llk}, \eqref{corre} is just the standard scattering amplitude in the massive case. 

An interesting problem is the scattering process with $m_1+n_1$ massless and $m_2+n_2$ massive particles. One should insert the massless fields at $\mathcal{I}^{\pm}$ and massive fields at $i^{\pm}$. A diagram with $m_1=n_1=m_2=n_2=1$ is shown in figure \ref{1111} for which one should define a correlator of mixed type 
\begin{align}
&\langle 0|\Sigma(u_4,\Omega_4)\varphi(y_3)\varphi^{(-)\dagger}(y_2)\Sigma^{(-)}(v_1,\Omega_1)|0\rangle\nn\\
&=\left(\frac{1}{8\pi^2\i}\times\frac{1}{2(2\pi)^{3/2}}\right)^2\int_0^\infty \d\omega_1\,\e^{\i\omega_1v_1}\int_0^\infty \d\omega_4\,\e^{-\i\omega_4u_4}\mathcal{A}_4(\bm p_1,\bm p_2,\bm p_3,\bm p_4),\label{6.3}
\end{align}  
where $\Sigma/\Sigma^{(-)}$ denotes the field at $\ci^+/\ci^-$ and $\mathcal A_4$ is the four-point scattering amplitude in the momentum space. For the massive fields, the momenta are related to the corresponding coordinates via \eqref{pmt} which should replaced by (2.35) of \cite{Liu:2024nfc} for massless fields. \eq{6.3} is a ``partial'' Carrollian amplitude since the integral transform is only applied to the massless fields. It is interesting to study this problem in the future.

\item \textbf{Non-linearity.} In our work, the essential part is the linear theory. It is crucial to include the non-linear parts to distinguish various massive theories. In massive spin 2 theory, one can find theories that are free from ghosts, including massive gravity from extra dimensions \cite{Dvali:2000hr}, new massive gravity in 3 dimensions \cite{Bergshoeff:2009hq}, and bi-gravity \cite{Hassan:2011zd} as well as multi-gravity \cite{Hinterbichler:2012cn}. They will lead to different holographic correlators on $H^3$. 
\end{itemize}

\acknowledgments 
The work of J.L. was supported by NSFC Grant No. 12005069. The work of W.-B. Liu is supported by “the Fundamental Research Funds for the Central Universities” with No. YCJJ20242112.

\appendix
\section{Massive fields in general dimensions}\label{secddim}
For the $d$-dimensional Minkowski spacetime, we still introduce $(\t,\r)$ as in \eqref{taurhodef} such that
\begin{align}
\d s^2=-\mathrm{d} \tau^2+\tau^2(\d \r^2+\sinh^2\r\d\Omega^2_{d-2})\equiv -\mathrm{d} \tau^2+\tau^2h_{ab}\mathrm{d}y^a\mathrm{d}y^b,
\end{align}
where the metric for unit sphere $S^{d-2}$ is still denoted by $\d\Omega^2_{d-2}=\g_{AB}\d x^A\d x^B$. Future timelike infinity $i^+$ is a unit $(d-1)$-dimensional hyperboloid $H^{d-1}$ with metric $h_{ab}$. The expressions for the Christoffel symbol \eqref{ksf} and Jacobi matrices \eqref{Jacobi} still hold. 

We directly consider the Proca field for simplicity. Using the saddle-point approximation to evaluate the mode expansion at large $\t$, we find
\begin{align}
A_\m=[\frac{m^{(d-3)/2}}{\t^{(d-1)/2}}\ca_\m(y)+O(\t^{-(d+1)/2})]\e^{-\i m\t}+\hc,
\end{align}
where the boundary field is defined through
\begin{align}
\ca_\m=\frac{1}{2(2\pi)^{(d-1)/2}}\sum_{\l}\epsilon_\m^\l(y) a_\l.\label{camexpddim}
\end{align}
The structure of the equation of motion is not changed, so we can still obtain the solution \eqref{solution}. Similarly, we derive the same fundamental commutator as \eqref{fundcom} and extended charges as \eqref{tfmxdefs1}. In consequence, there will also be an emerging spin charge as in \eqref{MXMY1} and we can define
\begin{align}
\cs_{s}&=-2\i m^2\int \d^{d-1}y\sqrt{h}\,s_{ab}\ca^{\dagger a}\ca^b.
\end{align}
The extended charge algebra is still \eqref{chargealgebrae}. 

If specialized to 3 dimensions, any 2-form on $H^2$ is proportional to $\ep_{ab}$ and we can rewrite $\cs_{s}$ as
\begin{align}
\cs_{s}&=-2\i m^2\int \d^2y\sqrt{h}\,s(y)\epsilon_{ab}\ca^{\dagger a}\ca^b.
\end{align}
Three of the commutators in \eqref{chargealgebrae} will be simplified to
\begin{subequations}\label{chargealgebra3d}
\begin{align}
[\cm_X,\cm_Y]&=\i\cm_{[X,Y]}+\i\cs_{o(X,Y)},\label{mxmys}\\
[\cm_X,\cs_s]&=\i\cs_{X(s)},\\
[\cs_{s_1},\cs_{s_2}]&=0,
\end{align}
\end{subequations}
where $o(X,Y)$ is now a function
\begin{align}
o(X,Y)=\frac{1}{2}\epsilon^{ab}o_{ab}(X,Y)=\ep^{ad}h^{bc}\nabla_{(a}X_{b)}\nabla_{(c}Y_{d)}.
\end{align}

\bibliographystyle{JHEP}
\bibliography{biblio}
\end{document}